\documentclass[reprint,twocolumn,superscriptaddress,showpacs, longbibliography]{revtex4-1}

\usepackage{amsmath,amssymb,amsfonts}
\usepackage{bm, latexsym}
\usepackage[dvipdfmx]{graphicx}
\usepackage{bbm,times}
\usepackage[normalem]{ulem}
\usepackage[usenames,dvipsnames]{color}

\begin{document}

%%\title{Correlation functions and finite-size scaling in a non-Hermitian XXZ spin chain}
\title{Universal properties of dissipative Tomonaga-Luttinger liquids: \\Case study of a non-Hermitian XXZ spin chain}
\author{Kazuki Yamamoto}
\email{yamamoto.kazuki.72n@st.kyoto-u.ac.jp}
\affiliation{Department of Physics, Kyoto University, Kyoto 606-8502, Japan}
\author{Masaya Nakagawa}
\affiliation{Department of Physics, University of Tokyo, 7-3-1 Hongo, Tokyo 113-0033, Japan}
\author{Masaki Tezuka}
\affiliation{Department of Physics, Kyoto University, Kyoto 606-8502, Japan}
\author{Masahito Ueda}
\affiliation{Department of Physics, University of Tokyo, 7-3-1 Hongo, Tokyo 113-0033, Japan}
\affiliation{RIKEN Center for Emergent Matter Science (CEMS), Wako, Saitama 351-0198, Japan}
\affiliation{Institute for Physics of Intelligence, University of Tokyo, 7-3-1 Hongo, Tokyo 113-0033, Japan}
\author{Norio Kawakami}
\affiliation{Department of Physics, Kyoto University, Kyoto 606-8502, Japan}

\date{\today}

\begin{abstract}
We demonstrate the universal properties of dissipative Tomonaga-Luttinger (TL) liquids by calculating correlation functions and performing finite-size scaling analysis of a non-Hermitian XXZ spin chain as a prototypical model in one-dimensional open quantum many-body systems. Our analytic calculation is based on effective field theory with bosonization, finite-size scaling approach in conformal field theory, and the Bethe-ansatz solution. Our numerical analysis is based on the density-matrix renormalization group generalized to non-Hermitian systems (NH-DMRG). We uncover that the model in the massless regime with weak dissipation belongs to the universality class characterized by the complex-valued TL parameter, which is related to the complex generalization of the $c=1$ conformal field theory. As the dissipation strength increases, the values of the TL parameter obtained by the NH-DMRG begin to deviate from those obtained by the Bethe-ansatz analysis, indicating that the model becomes massive for strong dissipation. Our results can be tested with the two-component Bose-Hubbard system of ultracold atoms subject to two-body loss.
\end{abstract}

\maketitle

%%%-----[Introduction]-----
\section{Introduction}
In recent years, open quantum systems have been actively studied both experimentally and theoretically, such as in driven-dissipative many-body systems \cite{Muller12, Daley14, Diehl16} and non-Hermitian (NH) quantum systems \cite{Ashida20}. In many cases, coupling to the environment causes decoherence of quantum states and it is often detrimental to their control. Remarkably, dissipation can also be instrumental in the preparation of novel states in open quantum systems. To date, a number of theoretical studies have shown that dissipation drastically alters various aspects of quantum many-body physics \cite{Diehl08NP, Kraus08, Witt08, Tomadin11, Boite13, Horstmann13, Buca20, Yamamoto20, Li20, Garcia09, Durr09, Rey14, Nakagawa20, Nakagawa21}. In particular, the interplay between unitary many-body dynamics and nonunitary state evolution due to dissipation leads to unconventional phase transitions \cite{Yamamoto19, Yamamoto21, Diehl10, Sieberer13, Hamazaki19, Hanai19, Matsumoto20, Lenke21}, quantum critical phenomena \cite{Ashida16, Ashida17, Sujit21, Kumar21, Nakagawa18, Louren18, Hanai20}, and measurement-induced entanglement transitions \cite{Fisher18, Smith19, Nahum19, Lunt20, Fuji20, Goto20, Tang20, Buchhold21, Muller21, Block21, Minato21}. Experimentally, high controllability of ultracold atoms has enabled observations of novel phases and phenomena unique to open quantum systems \cite{Ott13, Ott15, Ott16, Patil15, Schneider17, Gerbier20, Tomita19, Esslinger19, Konishi21}, e.g., continuous quantum Zeno effect \cite{Syassen08, Yan13, Tomita17}, loss-induced Dicke state \cite{Spon18}, and parity-time-symmetric NH quantum many-body systems \cite{Takasu20}. For the realization of NH quantum many-body systems, postselection of measurement outcomes by means of quantum-gas microscopy can be utilized \cite{Ashida20, Nakagawa20, Ashida16, Ashida17, Ott16Rev}. Such experimental progress has facilitated investigations of NH quantum systems.

In one-dimensional (1D) NH quantum many-body systems, one of the most intriguing phenomena is the dissipation-induced quantum criticality \cite{Dora20, Dora21, Moca21}. For example, the emergence of exceptional points accompanied by the divergence of the correlation length is reported \cite{Nakagawa21}, a quantum critical point in an interacting Bose gas is shifted by measurement backaction \cite{Ashida16}, and anomalous enhancement of the superfluid correlation occurs as a result of a semicircular renormalization-group flow \cite{Ashida17}. These studies have demonstrated that dissipation fundamentally alters the critical properties which have been studied in Hermitian quantum systems. Thus, a natural question arises about the universality of the unusual quantum critical phenomena. However, the universal properties of 1D NH quantum many-body systems are still elusive \cite{Wehefritz96, Wehefritz97, Couvreur17}.

One-dimensional quantum systems in equilibrium have widely been explored in condensed matter physics. Examples include various spin chains \cite{Haldane83} and the Hubbard model \cite{LiebWu68, Kawakami90A, Schulz90, Frahm90}. They show rich critical phenomena and universality emerging from quantum fluctuations in low-dimensional systems \cite{Haldane80, Haldane81C, Haldane81A, Haldane81}. Importantly, 1D strongly correlated systems realize the Tomonaga-Luttinger (TL) liquid, where the low-energy physics is described by massless collective modes \cite{Haldane80, Haldane81C, Haldane81A, Haldane81, Kawakami90, Kawakami91Condens, Kawakami91, Kawakami92}. A unified description of 1D quantum critical systems is given by the conformal field theory \cite{BPZ84, BPZ84stat, Daniel84, Brezin88, Cardy04, Francesco12}, where the TL liquid is characterized by the massless boson theory with the central charge $c=1$. For identifying the universality class of 1D critical systems, a useful fact is that the conformal dimensions are obtained from the energy gap due to a finite-size effect in the spectrum of the critical Hamiltonian. The efficient method to evaluate the conformal dimensions is finite-size scaling \cite{Cardy84a, Cardy84b, Cardy86, Cardy86log, Blote86, Affleck86, Hamer85, Hamer86, Woynarovich87, Francisco87}. However, it is highly nontrivial how to identify the unconventional universality class of NH quantum many-body systems from the finite-size scaling \cite{Wehefritz97}.

In this paper, we demonstrate the universal properties of dissipative TL liquids by calculating correlation functions and performing a finite-size scaling analysis of a NH XXZ spin chain. We first employ an effective field theory with bosonization to elucidate the long-distance properties of dissipative TL liquids, calculating two types of correlation functions called right-state correlation functions and biorthogonal correlation functions according to whether the right or left eigenstate is assigned to the bra vector in the expectation value. We then determine the parameters of the field theory from the exact Bethe-ansatz solution of the NH XXZ chain with the help of the finite-size scaling in conformal field theory. We find that the NH XXZ spin chain belongs to the universality class characterized by the complex-valued TL parameter $\tilde K$, which is related to the complex generalization of the $c=1$ conformal field theory. Finally, we give strong numerical evidence of the universal scaling with $\tilde K$ by calculating the energy spectrum and the correlation functions with the density-matrix renormalization group analysis generalized to non-Hermitian systems (NH-DMRG). The results show that the model is described by the nonunitary massless Gaussian theory for weak dissipation. On the other hand, when dissipation is increased, both the TL parameter and the velocity of excitations obtained by NH-DMRG start to deviate from those obtained by the Bethe-ansatz solution. This deviation indicates a significant finite-size effect and that the ground state can be gapped for strong dissipation. Our results can be tested with the two-component Bose-Hubbard system of ultracold atoms subject to two-body loss \cite{Tomita17}.

The rest of this paper is organized as follows. In Sec.~\ref{sec_model}, we derive the NH XXZ spin chain by starting from an experimentally relevant two-component Bose-Hubbard model with two-body loss and applying a quantum trajectory method to the Lindblad master equation. We then apply the bosonization method to obtain the NH sine-Gordon model in which the coefficients of the Gaussian term and the cosine term become complex-valued. The analytical calculations in the subsequent two sections are conducted in the massless regime. We give correlation functions in Sec.~\ref{sec_correlation} by using the effective field theory. Section~\ref{sec_ExactFinite} is devoted to the exact solution of the NH XXZ model by the Bethe ansatz method. We obtain the energy spectrum in a finite system and demonstrate that it is consistent with the finite-size scaling formula generalized to NH TL liquids. We also discuss stability conditions of the NH TL liquids. In Sec.~\ref{sec_NH-DMRG}, we report the NH-DMRG results for the NH XXZ spin chain which are obtained without the assumption that the system is in the massless regime. We also compare them with the corresponding analytical results obtained in Secs.~\ref{sec_correlation} and \ref{sec_ExactFinite}. We finally summarize the results and discuss some outlooks in Sec.~\ref{sec_discussion}.
%%In Appendix~\ref{sec_bosonization}, we explain the detailed calculation of the bosonization for the NH XXZ model. In Appendix~\ref{sec_algorithm}, we give a summary of the NH-DMRG algorithm. Appendix~\ref{sec_gappedNH-DMRG} provides a report of the problem of NH-DMRG in the gapped regime for those who are going to be engaged in NH-DMRG.

%%%-----[model]------
\section{Model}
\label{sec_model}
In this section, we first derive the NH XXZ model as an effective model of a two-component Bose-Hubbard system subject to two-body loss. Then, we bosonize the Hamiltonian to analyze the TL-liquid properties, and obtain an effective TL Hamiltonian.
\subsection{Non-Hermitian XXZ model}
NH spin models in the presence of dissipation have been proposed as prototypical dissipative quantum systems relevant to experiments in ultracold atoms \cite{Lee14, Shibata19, Nakagawa20, Buca20}. However, NH many-body phenomena in dissipative spin systems are less explored \cite{Shibata19, Nakagawa20, Buca20}, compared with the other dissipative spin models that can be mapped to noninteracting NH systems or to those in the master equation frameworks \cite{Prosen08, Lee13, Joshi13, Lee14}. Here we follow Ref.~\cite{Nakagawa20} to derive a NH XXZ spin chain from a dissipative two-component Bose-Hubbard model of ultracold atoms. The unitary dynamics of the system without loss is governed by the two-component Bose-Hubbard model
\begin{align}
H=&-t_h\sum_{j,\sigma=\uparrow,\downarrow}(b_{j+1\sigma}^\dag b_{j\sigma}+\mathrm{H.c.})+\sum_j U_{\uparrow\downarrow}n_{j\uparrow}n_{j\downarrow}\notag\\
&+\sum_{j,\sigma} \frac{U_{\sigma\sigma}}{2}n_{j\sigma}(n_{j\sigma}-1),
\label{eq_BH}
\end{align}
where $b_{j\sigma}$ is the annihilation operator of a boson with spin $\sigma$ at site $j$, $n_{j\sigma}=b_{j\sigma}^\dag b_{j\sigma}$, and $t_h>0$ is the hopping amplitude, which we assume to be the same for every site and spin state. We assume that the on-site interaction is repulsive: $U_{\sigma\sigma'}>0$. When the system is subject to two-body particle loss, the dynamics is described by the Lindblad master equation \cite{Lindblad76}
\begin{align}
\frac{d\rho}{dt}
&=-i[H,\rho]-\frac{1}{2}\sum_{j\sigma\sigma^\prime}(\{L_{j\sigma\sigma^\prime}^\dagger{L}_{j\sigma\sigma^\prime},\rho\}-2L_{j\sigma\sigma^\prime}\rho{L}_{j\sigma\sigma^\prime}^\dagger)\notag\displaybreak[2]\\
&=-i(H_{\mathrm{eff}}\rho-\rho{H}_{\mathrm{eff}}^\dagger)+\sum_{j\sigma\sigma^\prime}{L}_{j\sigma\sigma^\prime}\rho{L}_{j\sigma\sigma^\prime}^\dagger,
\label{eq_lindblad}
\end{align}
where $\rho$ is the density matrix of the system, $L_{j\sigma\sigma^\prime}=\sqrt{\gamma_{\sigma\sigma^\prime}}b_{j\sigma}b_{j\sigma^\prime}$ is the Lindblad operator that describes two-body loss with rate $\gamma_{\sigma\sigma^\prime}>0$ \cite{Syassen08, Yan13, Rey14, Garcia09, Durr09, Tomita17, Tomita19, Spon18, Yamamoto19, Yamamoto21, Nakagawa20, Nakagawa21, He20, Xu20, Yoshida20}. In this case, the effective Hamiltonian $H_{\mathrm{eff}}$ is given by
\begin{align}
H_{\mathrm{eff}}=&-t_h\sum_{j\sigma}(b_{j+1\sigma}^\dag b_{j\sigma}+\mathrm{H.c.})+\sum_j (U_{\uparrow\downarrow}-i\gamma_{\uparrow\downarrow})n_{j\uparrow}n_{j\downarrow}\notag\\
&+\sum_{j\sigma} \frac{U_{\sigma\sigma}-i\gamma_{\sigma\sigma}}{2}n_{j\sigma}(n_{j\sigma}-1),
\label{eq_BHeff}
\end{align}
where we have used $H_{\mathrm{eff}}=H-\frac{i}{2}\sum_{j\sigma\sigma^\prime}L_{j\sigma\sigma^\prime}^\dagger{L}_{j\sigma\sigma^\prime}$ \cite{Daley14} and $\gamma_{\uparrow\downarrow}=\gamma_{\downarrow\uparrow}$. In ultracold atoms, the approximations involved in deriving the Lindblad master equation are typically satisfied to sufficient precision \cite{Daley14}. We note that Eq.~\eqref{eq_BHeff} gives a negative imaginary part of the energy and its eigenstates decay due to atom loss. As discussed below, we consider the longest-surviving state which is given by the largest imaginary part of the energy (smallest absolute value of the negative imaginary part of the energy) in Eq.~\eqref{eq_BHeff}.

We invoke unraveling of the dynamics of the density matrix into quantum trajectories \cite{Daley14}, each of which obeys the Schr\"{o}dinger evolution with the effective Hamiltonian $H_\mathrm{eff}$ interrupted by quantum jumps described by the jump operators $L_{j\sigma\sigma'}$. We consider a strongly correlated regime $U_{\sigma\sigma^\prime}\gg t_h$ and assume that each site is occupied on average by one particle so that a Mott insulating state is realized as an initial state. For simplicity, we assume $U_{\uparrow\uparrow}=U_{\downarrow\downarrow}\equiv U$ and $\gamma_{\uparrow\uparrow}=\gamma_{\downarrow\downarrow}\equiv \gamma$. Then, the second-order perturbation theory with respect to $t_h$ reduces the effective NH Hamiltonian \eqref{eq_BHeff} to the NH XXZ model \cite{Nakagawa20, Duan03}
\begin{align}
H_{\mathrm{eff}}=&(J_{\mathrm{eff}}^\perp + i\Gamma^\perp)\sum_j (S_{j+1}^xS_j^x + S_{j+1}^yS_j^y)\notag\\&+(J_{\mathrm{eff}}^z+i\Gamma^z) \sum_j S_{j+1}^zS_j^z\notag\displaybreak[2]\\
=&(J_{\mathrm{eff}}^\perp + i\Gamma^\perp)\sum_j(S_{j+1}^xS_j^x + S_{j+1}^yS_j^y+\Delta_\gamma S_{j+1}^zS_j^z)
\label{eq_Hferroeff},
\end{align}
where $S_j^\alpha$ ($\alpha=x$, $y$, $z$) are the spin-$1/2$ operators, $J_{\mathrm{eff}}^\perp=-4t_h^2U_{\uparrow\downarrow}/(U_{\uparrow\downarrow}^2+\gamma_{\uparrow\downarrow}^2)$, $\Gamma^\perp=-4t_h^2\gamma_{\uparrow\downarrow}/(U_{\uparrow\downarrow}^2+\gamma_{\uparrow\downarrow}^2)$, $J_{\mathrm{eff}}^z=-J_{\mathrm{eff}}^\perp-8t_h^2U/(U^2+\gamma^2)$, $\Gamma^z=-\Gamma^\perp-8t_h^2\gamma/(U^2+\gamma^2)$, and we have ignored a constant term. We note that, since the energy spectrum of the original effective Hamiltonian \eqref{eq_BHeff} always has a negative imaginary part, the imaginary part of the energy of the NH XXZ model \eqref{eq_Hferroeff} should be negative if we include the ignored constant term. As the constant term does not change the structure of the energy spectrum, the eigenstate with the smallest decay rate of the effective Hamiltonian \eqref{eq_Hferroeff} is given by the one with the energy having the largest imaginary part, which corresponds to the smallest absolute value of the negative imaginary part of the energy in the original effective Hamiltonian \eqref{eq_BHeff}. Here, the anisotropy parameter
\begin{align}
\Delta_\gamma &\equiv \frac{J_{\mathrm{eff}}^z+i\Gamma^z}{J_{\mathrm{eff}}^\perp + i\Gamma^\perp}\notag\\
&=2\frac{U_{\uparrow\downarrow}U+\gamma_{\uparrow\downarrow}\gamma}{U^2+\gamma^2}-1 + 2i\frac{U_{\uparrow\downarrow}\gamma-U\gamma_{\uparrow\downarrow}}{U^2+\gamma^2},
\end{align}
becomes complex in general, and its imaginary part can be either positive or negative. However, in the case where $U=\gamma\equiv G$ and $U_{\uparrow\downarrow}=\gamma_{\uparrow\downarrow}\equiv G_{\uparrow\downarrow}$, the anisotropy parameter becomes $\Delta_\gamma=2G_{\uparrow\downarrow}/G-1 \in \mathbb{R}$. Thus, we have to pay attention to the fact that the anisotropy parameter $\Delta_\gamma$ can be real even in the presence of dissipation.

In the case of $\Delta_\gamma\in\mathbb R$, the complex coefficients only appear as an overall constant and the eigenstates are the same as those of the Hermitian XXZ model \cite{Nakagawa20}. As $J_{\mathrm{eff}}^\perp$ and $\Gamma^\perp$ have the same sign ($J_{\mathrm{eff}}^\perp,\Gamma^\perp<0$), higher energy states (i.e., states with a larger real part of the energy) have smaller decay rates. Thus, after a sufficiently long time, only the high-energy spin states can survive in the Schr\"{o}dinger evolution under $H_{\mathrm{eff}}$. In particular, the system approaches the ground state of the following XXZ model in the long-time limit:
\begin{align}
H_{\mathrm{eff}}^\mathrm{XXZ}&= \frac{J}{2}\sum_j (S_{j+1}^+S_j^- + S_{j+1}^-S_j^+) + J\Delta_\gamma \sum_j S_{j+1}^zS_j^z
\label{eq_Heff},
\end{align}
where $S_j^\pm = S_j^x \pm iS_j^y$ and $J>0$. Note that $\mathrm{sgn}[J]=-\mathrm{sgn}[J_{\mathrm{eff}}^\perp]$.

When $\Delta_\gamma\in\mathbb{C}$, the eigenstates of the effective NH Hamiltonian \eqref{eq_Hferroeff} are not equivalent to those of the Hermitian Hamiltonian and therefore unusual quantum critical phenomena may take place due to non-Hermiticity. In this case, the longest-surviving state with the largest imaginary part of the eigenenergy of the effective NH Hamiltonian \eqref{eq_Hferroeff} is given by the ground state with the lowest real part of the eigenenergy in the NH XXZ model \eqref{eq_Heff}, if the imaginary part of $\Delta_\gamma$ is sufficiently small so that no level crossing occurs. Therefore, for convenience of calculations, we hereafter focus on the ground state of the NH XXZ model \eqref{eq_Heff}.

The dynamics under the NH effective Hamiltonian \eqref{eq_Hferroeff} is realized when we measure the particle number with quantum-gas microscopy and postselect measurement outcomes in which the particle number is equal to that of the initial state. In Ref.~\cite{Nakagawa20}, it was shown that the highest energy state in Eq.~\eqref{eq_Hferroeff} is obtained even in quantum trajectories which involve quantum jumps, if the spin correlation is measured only at the sites that are occupied by single particles. This is a consequence of the spin-charge separation in 1D systems \cite{Giamarchi03}, where the spin degrees of freedom are decoupled from holes created by quantum jumps. Thus, by postselecting the occupied sites with quantum-gas microscopy, the spin correlations after a sufficiently long time are expected to be described by those of the ground state of Eq.~\eqref{eq_Heff}. Here, we note that the steady state of the Lindblad master equation \eqref{eq_lindblad} is the vacuum. However, as the dynamics under the NH effective Hamiltonian \eqref{eq_Hferroeff} is obtained by postselecting special measurement outcomes in which there are no loss events, it realizes a nontrivial steady state, which is different from the state obtained under the Lindblad master equation \eqref{eq_lindblad} \cite{Nakagawa20}. In this paper, we analyze the NH XXZ model \eqref{eq_Heff} and elucidate how unconventional universal properties of TL liquids emerge in NH spin chains.

\subsection{Non-Hermitian Tomonaga-Luttinger model}
In order to elucidate the long-distance behavior of dissipative TL liquids, we bosonize the NH XXZ Hamiltonian \eqref{eq_Heff}. We use the standard boson mapping as detailed in Appendix~\ref{sec_bosonization} \cite{Giamarchi03}, and take the continuum limit by introducing $S^+(x)=S_j^+/\sqrt{a}$, $S^z(x)=S_j^z/a$, where $a$ is the lattice spacing. After the bosonization procedure, the spin operators in the continuum limit are written as
\begin{align}
S^z(x)&=-\frac{1}{\pi}\nabla\phi(x)+\frac{(-1)^x}{\pi\alpha}\cos(2\phi(x)),\label{eq_Sz}\\
S^+(x)&=\frac{e^{-i\theta(x)}}{\sqrt{2\pi\alpha}}\left((-1)^x + \cos(2\phi(x))\right),\label{eq_S+}
\end{align}
where $x$ is related to the lattice coordinate as $x=aj$ with $a=1$, $\alpha$ is a cutoff, and the bosonic fields $\phi(x)$ and $\theta(x)$ satisfy the commutation relation $[\phi(x_1), \nabla \theta(x_2)]=i\pi\delta(x_2-x_1)$. Then, we obtain the NH sine-Gordon Hamiltonian
\begin{align}
H_{\mathrm{eff}}^\mathrm{sG}=H_{\mathrm{eff}}^{\mathrm{TL}}- \frac{2\tilde g_3}{(2\pi\alpha)^2}\int dx \cos(4\phi(x)),
\label{eq_NHSG}
\end{align}
where $\tilde g_3$ is a complex-valued coefficient that depends on $\Delta_\gamma$, and
\begin{align}
H_{\mathrm{eff}}^{\mathrm{TL}}=\frac{1}{2\pi}\int dx \Big[\tilde u\tilde K(\nabla\theta(x))^2 + \frac{\tilde u}{\tilde K}(\nabla\phi(x))^2\Big].
\label{eq_Heff0}
\end{align}
Here, $\tilde K$ is the complex-valued TL parameter and $\tilde u$ is the complex-valued velocity of excitations. We obtain the exact solutions of $\tilde K$ and $\tilde u$ by using the Bethe-ansatz method generalized to NH systems in Sec.~\ref{sec_ExactFinite}. Here and henceforth, we use the symbol $\tilde A$ with a tilde to emphasize that a quantity $\tilde A$ is complex. In Sec.~\ref{sec_correlation}, we consider the situation where the model is in the massless regime, and we analyze the NH TL model $H_{\mathrm{eff}}^{\mathrm{TL}}$.

%%%-----[result]------
\section{Correlation functions}
\label{sec_correlation}
In NH systems, a right eigenstate, which is defined by $H_\mathrm{eff}^\mathrm{TL}|\Psi^R\rangle=E|\Psi^R\rangle$, and a left eigenstate, which is defined by $H_\mathrm{eff}^\mathrm{TL\dag}|\Psi^L\rangle=E^*|\Psi^L\rangle$, are different from each other. Therefore, two types of correlation functions can emerge according to whether the right or left eigenstate is assigned to the bra vector in the expectation value. The first type is defined by ${}_L\langle\cdots\rangle_R\equiv \langle\Psi_0^L|\cdots|\Psi_0^R\rangle/\langle\Psi_0^L|\Psi_0^R\rangle$, where $|\Psi_0^L\rangle$ and $|\Psi_0^R\rangle$ are the left and right ground states (in the sense of the real part of the energy) of $H_\mathrm{eff}^\mathrm{TL}$, respectively. This type of correlation functions is calculated through path integrals \cite{Yamamoto19}. The second type is defined by ${}_R\langle\cdots\rangle_R\equiv \langle\Psi_0^R|\cdots|\Psi_0^R\rangle/\langle\Psi_0^R|\Psi_0^R\rangle$, which is calculated by the wave functional approach \cite{Ashida16, Furukawa11}. Here, we note that the subscripts $L$ and $R$ for the bra and ket vectors stand for the left and right eigenstates of the NH Hamiltonian \eqref{eq_Heff0}, and they are not related to the left and right branches of the TL model. We call the correlation functions ${}_L\langle\cdots\rangle_R$ and ${}_R\langle\cdots\rangle_R$  the biorthogonal correlation function and the right-state correlation function, respectively.

In the postselected sector with no loss events, the dynamics of the system is described by the Schr\"{o}dinger equation $i\partial_t|\psi\rangle = H_\mathrm{eff}|\psi\rangle$, which gives the right ground state of the NH XXZ Hamiltonian (6) in the long-time limit. Then, the right-state correlation function is obtained as a standard quantum-mechanical expectation value for the state $|\psi(t\to\infty)\rangle=|\Psi_0^R\rangle$ and corresponds to an experimentally measured physical quantity. On the other hand, the biorthogonal correlation function gives a natural extension of the correlation function that can be calculated with a field-theoretical method in NH systems. We also emphasize that the biorthogonal correlation functions are directly related to the complex extension of the $c=1$ conformal field theory as detailed in Sec.~\ref{sec_ExactFinite}. We calculate both correlation functions in this section.

\subsection{Biorthogonal correlation functions}
In this subsection, we discuss the biorthogonal correlation functions. We first calculate the correlation functions of the fields $\phi$ and $\theta$, and use them to obtain the correlation functions of the spin operators. An important point is that a convergence problem of the Gaussian integration occurs due to the complex nature of $\tilde u$ and $\tilde K$. Our calculation is based on the path integral formalism \cite{Yamamoto19}.

\subsubsection{Path-integral formalism and correlation functions of $\phi$ and $\theta$}
We start with the partition function defined by
\begin{align}
Z&=\mathrm{Tr}[e^{-\beta H_{\mathrm{eff}}^{\mathrm{TL}}}]=\int\mathcal D \phi \mathcal D \Pi e^{-S},\\
S&=-\int_0^\beta d\tau\int_{-\infty}^\infty dx[i\Pi\partial_\tau\phi-H_{\mathrm{eff}}^{\mathrm{TL}}(\phi, \Pi)],
\end{align}
where $\Pi(x, \tau) = \nabla\theta(x, \tau) / \pi$. We note that, as temperature is not well defined in generic open quantum systems, we only consider the limit of infinite $\beta$ to elucidate the physics of the ground state, which is defined by the eigenstate that has the lowest real part of the eigenspectrum \cite{Ashida16, Yamamoto19}. Thus, $\beta$ is a parameter used to formulate a path integral and should not be regarded as the temperature of the system. In the following, we calculate the equal-time correlation function ${}_L\langle [\phi(x_1, 0)-\phi(x_2, 0)]^2\rangle_R$. It is rewritten by using the Fourier transformation $\phi(x, \tau)=\frac{1}{\beta L}\sum_{\bm q}e^{i(xq-\omega_n\tau)}\phi(\bm q)$ as
\begin{align}
{}_L\langle &[\phi(x_1, 0)-\phi(x_2, 0)]^2\rangle_R
=\frac{1}{(\beta L)^2}\sum_{\bm q_1, \bm q_2}\notag\\
&\times\left[{}_L\langle\phi(\bm q_1)\phi(\bm q_2)\rangle_R(e^{ik_1 x_1}-e^{ik_1 x_2})(e^{ik_2 x_1}-e^{ik_2 x_2})\right],
\end{align}
where $\bm q = (k, \omega_n / \tilde u)$, $\omega_n = 2\pi n / \beta$ is the Matsubara frequency of bosons, and $L$ is the length of the lattice. The action is calculated as
\begin{align}
S =&\int_0^\beta d\tau\int_{-\infty}^\infty dx\Bigg[-i\frac{1}{\pi}\nabla\theta(x, \tau)\partial_\tau\phi(x, \tau)\notag\\
&+\frac{1}{2\pi}\left(\tilde u \tilde K(\nabla \theta(x, \tau))^2+\frac{\tilde u}{\tilde K}(\nabla \phi(x,\tau))^2\right)\Bigg]\notag\\
=&\frac{1}{2\beta L}\sum_{\bm q}(\theta^*(\bm q),\phi^*(\bm q))M
\left(\begin{matrix}
\theta(\bm q)\\
\phi(\bm q)
\end{matrix}
\right),
\label{eq_action_M}
\end{align}
where the matrix $M$ is given by
\begin{align}
M=
\left(
\begin{matrix}
k^2\frac{\tilde u \tilde K}{\pi}&\frac{ik\omega_n}{\pi}\\
\frac{ik\omega_n}{\pi}&k^2\frac{\tilde u}{\tilde K\pi}
\end{matrix}
\right).
\end{align}
In the Hermitian case, the Gaussian integration with the action \eqref{eq_action_M} always converges. However, in the NH case, the Gaussian integration can be divergent because the velocity $\tilde u$ of excitations and the TL parameter $\tilde K$ become complex. To ensure the convergence of the Gaussian integration, the Hermitian part of the matrix $M$ should be positive definite, i.e., $\mathrm{Re} (\tilde u \tilde K)>0$ and $\mathrm{Re}(\tilde u/\tilde K)>0$. These conditions are equivalent to those that ensure the energy spectrum of the NH TL liquids to be bounded from below (see Sec.~\ref{sec_stability}). Thus, if these conditions are not satisfied, the NH TL liquids become unstable.
With these conditions, we can conduct further calculations and obtain
\begin{align}
{}_L\langle\phi(\bm q_1)\phi(\bm q_2)\rangle_R&=\frac{1}{Z_\phi}\int\mathcal D\phi e^{-S_\phi}\phi(\bm q_1)\phi(\bm q_2)\notag\\
&= \frac{\pi \tilde K \delta_{\bm q_1, -\bm q_2}L\beta}{\frac{\omega_{n}^2}{\tilde u} + \tilde uk_1^2},
\end{align}
where
\begin{gather}
Z_\phi = \int\mathcal D \phi e^{-S_\phi},\\
S_\phi=\frac{1}{\beta L}\sum_{\bm q}\frac{1}{2\pi \tilde K}\left(\frac{\omega_n^2}{\tilde u} + \tilde uk^2\right)\phi^*(\bm q)\phi(\bm q).
\end{gather}
Then, we arrive at
\begin{align}
{}_L\langle [\phi(x, 0)-\phi(0, 0)]^2\rangle_R&=\frac{1}{\beta L}\sum_{\bm q}\frac{\pi \tilde K }{\frac{\omega_n^2}{\tilde u} + \tilde uk^2}(2-2\cos kx)\notag\\
&=\frac{\tilde K}{2}\log\left(\frac{x^2+\alpha^2}{\alpha^2}\right),
\label{eq_phiLR}
\end{align}
where we have introduced a cutoff $\alpha$ and taken the limit $\beta\to\infty$. We note that the limit $\beta\to\infty$ is safely taken under the condition $\mathrm{Re} (\tilde u \tilde K)>0$ and $\mathrm{Re}(\tilde u/\tilde K)>0$. We can apply a similar procedure to ${}_L\langle [\theta(x, 0)-\theta(0, 0)]^2\rangle_R$ by replacing $\tilde K$ with $1 / \tilde K$, and obtain
\begin{align}
{}_L\langle [\theta(x, 0)-\theta(0, 0)]^2\rangle_R=\frac{1}{2\tilde K}\log\left(\frac{x^2+\alpha^2}{\alpha^2}\right).
\label{eq_thetaLR}
\end{align}
Finally, by using the formula 
\begin{align}
&{}_L\Big\langle \exp[i\sum_j(A_j\phi(x_j,0)+B_j\theta(x_j,0))]\Big\rangle_R\notag\\
&=\exp\Big[-\frac{1}{2}{}_L\big\langle[\sum_j(A_j\phi(x_j,0)+B_j\theta(x_j,0))]^2\big\rangle_R\Big],
\end{align}
and Eqs.~\eqref{eq_phiLR} and \eqref{eq_thetaLR}, we arrive at
\begin{gather}
{}_L\langle e^{i(2\phi(x, 0)-2\phi(0, 0))}\rangle_R=\left(\frac{\alpha}{x}\right)^{2\tilde K},\label{eq_expphiLR}\\
{}_L\langle e^{i(2\theta(x, 0)-2\theta(0, 0))}\rangle_R=\left(\frac{\alpha}{x}\right)^{\frac{2}{\tilde K}}.\label{eq_expthetaLR}
\end{gather}
We see that the correlation functions of the fields $\phi$ and $\theta$ are characterized by the complex-valued TL parameter $\tilde K$. This fact is related to the complex generalization of the $c=1$ conformal field theory, which is discussed in Sec.~\ref{sec_ExactFinite}.

\subsubsection{Correlation function of the spin operators}
We here calculate the correlation functions of the spin operators defined in Eqs.~\eqref{eq_Sz} and \eqref{eq_S+}. For the correlation function of $S^z$, we obtain the free-fermion-like correlation $1/x^2$ and the power law characterized by $\tilde K$ by performing the Gaussian integration with Eq.~\eqref{eq_expphiLR} as
\begin{align}
{}_L\langle S^z(x, 0)S^z(0, 0)\rangle_R
=&\frac{1}{\pi^2}{}_L\langle \nabla\phi(x,0)\nabla\phi(0,0)\rangle_R\notag\\
+ \frac{(-1)^x}{(2\pi\alpha)^2}&{}_L\langle e^{i2(\phi(x,0)-\phi(0,0))}+\mathrm{H.c.}\rangle_R\notag\\
= -\frac{\tilde K}{2\pi^2}\frac{1}{x^2} &+ \tilde C_2(-1)^x\left(\frac{1}{x}\right)^{2\tilde K},
\end{align}
where $\tilde C_2$ is a complex-valued nonuniversal amplitude. Here, we note that $x$ is defined on the lattice $x=aj$, where we set $a=1$. For the $x$-$y$ component of the correlation function, we obtain the power law characterized by $1/\tilde K$ and the linear combination of $\tilde K$ and $1/\tilde K$ by using Eqs.~\eqref{eq_expphiLR} and \eqref{eq_expthetaLR} as
\begin{align}
{}_L\langle S^+(x, 0)S^-(0, 0)\rangle_R=&\frac{1}{2\pi\alpha}{}_L\langle e^{-i(\theta(x,0)-\theta(0,0))}[(-1)^x\notag\\
&+ \frac{1}{4}(e^{2i(\phi(x,0)-\phi(0,0))}+\mathrm{H.c.})]\rangle_R\notag\\
=\tilde C_3\left(\frac{1}{x}\right)^{2\tilde K+\frac{1}{2\tilde K}}&+\tilde C_4(-1)^x\left(\frac{1}{x}\right)^{\frac{1}{2\tilde K}},
\end{align}
where the off-diagonal terms with respect to the fields $\phi$ and $\theta$ such as ${}_L\langle\theta(x, 0)\phi(0,0)\rangle_R$ only contribute to the sign, and we have used the fact that ${}_L\langle \exp[i\sum_j(A_j\phi(x_j,0)+B_j\theta(x_j,0))]\rangle_R=0$ when $\sum_iA_i\neq0$ or $\sum_iB_i\neq0$. Here, $\tilde C_3$ and $\tilde C_4$ are nonuniversal complex-valued amplitudes, which satisfy $\tilde C_3<0$ in the Hermitian limit. Thus, we see that the power law decay of the biorthogonal correlation functions of the spin operators is universally characterized by the complex-valued TL parameter $\tilde K$. 

\subsection{Right-state correlation functions}
In this subsection, we study the right-state correlation functions. We first calculate the correlation functions of the fields $\phi$ and $\theta$, and then obtain the correlation functions of the spin operators by using them. Our calculation is based on the generalization of the harmonic oscillator to NH systems \cite{Ashida16, Furukawa11}.

\subsubsection{Ground-state wave function and correlation functions of $\phi$ and $\theta$}
We start with the NH TL Hamiltonian \eqref{eq_Heff0}, which is rewritten as
\begin{align}
H_{\mathrm {eff}}^{\mathrm{TL}}=\frac{1}{2\pi}\int dx\left[v_Je^{-i\delta_J}(\nabla\theta(x))^2+v_Ne^{-i\delta_N}(\nabla\phi(x))^2\right],\label{eq_Heff0v}
\end{align}
where $v_J>0$, $v_N>0$, and $\delta_J, \delta_N \in \mathbb R$.

The case with $\delta_J=0$ was analyzed in Ref.~\cite{Ashida16}. Since $e^{i\delta_J}H_\mathrm{eff}^\mathrm{TL}$ reduces to this case, the energy eigenspectrum and the ground state in the present case can be obtained by a straightforward extension of the results in Ref.~\cite{Ashida16}. The wave function of the ground state is given by
\begin{align}
\langle\{\phi_k\}|\Psi_0^R\rangle=\frac{1}{\sqrt {\mathcal N}} \exp\left(-\frac{e^{-i(\delta_N-\delta_J)/2}}{K^\prime}\sum_{k>0}k|\phi_k|^2\right),
\label{eq_wavefunharmonic}
\end{align}
where $\mathcal N$ is a normalization constant, $K^\prime=\sqrt{v_J/v_N}\in \mathbb R$, and $|\{\phi_k\}\rangle$ is an eigenstate of $\phi(x)$ defined by
\begin{gather}
\phi(x)|\{\phi_k\}\rangle=\sqrt{\frac{\pi}{L}}\sum_{k>0}(\phi_k e^{ikx}+\phi_k^* e^{-ikx})|\{\phi_k\}\label{eq_phik}\rangle.
\end{gather}
The field $\theta(x)$ induces the shift of the eigenstate $|\{\phi_k\}\rangle$ as follows:
\begin{align}
e^{2i\theta(x)}|\{\phi_k\}\rangle
=&\left|\left\{\phi_k-\frac{2i}{k}\sqrt{\frac{\pi}{L}}e^{-ikx}\right\}\right\rangle.
\end{align}
The eigenenergies are given by
\begin{align}
E =v^\prime e^{-i(\delta_N+\delta_J)/2}\sum_{k>0}k (n_k^+ +n_k^- +1),
\label{eq_energyharmonic}
\end{align} 
where $v^\prime=\sqrt{v_Jv_N}$, and $n_k^+$ and $n_k^-$ are nonnegative integers. By using the ground-state wave function \eqref{eq_wavefunharmonic}, the correlation functions of the fields $\phi$ and $\theta$ are calculated as \cite{Ashida16}
\begin{align}
&{}_R\langle e^{2i\phi(x)}e^{-2i\phi(0)}\rangle_R=\left(\frac{\alpha}{x}\right)^{2K_\phi},\label{eq_expphiRR}\\
&{}_R\langle e^{2i\theta(x)}e^{-2i\theta(0)}\rangle_R=\left(\frac{\alpha}{x}\right)^{\frac{2}{K_\theta}},\label{eq_expthetaRR}
\end{align}
where a cutoff $\alpha$ is introduced, and the critical exponents are defined by $K_\phi=K^\prime/\cos((\delta_N-\delta_J)/2)$ and $K_\theta=K^\prime \cos((\delta_N-\delta_J)/2)$. 

Here, we have to pay attention to the fact that the wave function \eqref{eq_wavefunharmonic} should be normalizable and the real part of the eigenvalues \eqref{eq_energyharmonic} should be bounded from below. For the model in Ref.~\cite{Ashida16}, these conditions are equivalent to each other. However, in our model, these two conditions are inequivalent due to the phase factor $e^{-i\delta_J}$ in the eigenspectrum \eqref{eq_energyharmonic}. In order that the wave function \eqref{eq_wavefunharmonic} and that of the zero mode can be normalized, $\delta_N$ and $\delta_J$ should satisfy $-\pi/2<\delta_N-\delta_J<\pi/2$. For the eigenenergies, in order for the real part of the eigenspectrum including the zero-mode contribution to be bounded from below, $\delta_N$ and $\delta_J$ should satisfy $-\pi/2<\delta_N<\pi/2$ and $-\pi/2<\delta_J<\pi/2$. Further discussions on the condition for realizing NH TL liquids are given in Sec.~\ref{sec_ExactFinite}.

Importantly, the right-state correlation functions of the NH TL liquid are characterized by the two critical exponents $K_\phi$ and $K_\theta$, though they coincide in the Hermitian limit \cite{Ashida16}. Since we can rewrite the complex-valued TL parameter as $\tilde K = K^\prime e^{i(\delta_N-\delta_J)/2}$ by comparing Eqs.~\eqref{eq_Heff0} and \eqref{eq_Heff0v}, the two critical exponents $K_\phi=K^\prime/\cos((\delta_N-\delta_J)/2)$ and $K_\theta=K^\prime \cos((\delta_N-\delta_J)/2)$ can be compactly written down as
\begin{align}
\frac{1}{K_\phi}=\mathrm{Re}\frac{1}{\tilde K},\label{eq_Kphi}\\
K_\theta=\mathrm{Re}\tilde K.\label{eq_Ktheta}
\end{align}
We see that the critical exponents for the fields $\phi$ and $\theta$ are related to each other through Eqs.~\eqref{eq_Kphi} and \eqref{eq_Ktheta}, and both are defined by the complex-valued TL parameter $\tilde K$. Thus, we conclude that all the universal properties of the biorthogonal and right-state correlation functions are encoded in the complex-valued TL parameter $\tilde K$, which is related to the NH generalization of the $c=1$ conformal field theory as discussed in Sec.~\ref{sec_ExactFinite}.

\subsubsection{Correlation functions of the spin operators}
We now calculate the correlation functions of the spin operators \eqref{eq_Sz} and \eqref{eq_S+}. By performing the Gaussian integration and using Eq.~\eqref{eq_expphiRR}, we obtain the correlation function of $S^z$ as
\begin{align}
{}_R\langle S^z(x)S^z(0)\rangle_R&=\frac{1}{\pi^2}{}_R\langle \nabla\phi(x)\nabla\phi(0)\rangle_R\notag\\
&\quad+ \frac{(-1)^x}{(2\pi\alpha)^2}{}_R\langle e^{i2\phi(x)-i2\phi(0)}+\mathrm{H.c.}\rangle_R\notag\\
&= -\frac{K_\phi}{2\pi^2}\frac{1}{x^2} + C_2(-1)^x\left(\frac{1}{x}\right)^{2K_\phi},\label{eq_SzRR}
\end{align}
where $C_2$ is a nonuniversal real correlation amplitude. We note that $x=aj$ is defined on the lattice and that we set $a=1$. We see that the correlation function of $S^z$ has the free-fermion-like correlation $1/x^2$ and anomalous power-law decay characterized by $K_\phi$. Next, we go on to calculate the $x$-$y$ component of the correlation function. The $x$-$y$ component of the correlation function is rewritten as
\begin{align}
{}_R\langle S^+(x)S^-(0)\rangle_R = &\frac{1}{2\pi\alpha}{}_R\langle e^{-i(\theta(x)-\theta(0))}[(-1)^x\notag\\
&+ \frac{1}{4}(e^{2i(\phi(x)-\phi(0))}+\mathrm{H.c.})]\rangle_R.
\end{align}
To proceed further with calculations, we consider the condition in which a general expectation value,
\begin{align}
{}_R\langle e^{-i(\theta(x)-\theta(0))}e^{i(A\phi(x)+B\phi(0))}\rangle_R,
\label{eq_corr_general}
\end{align}
becomes nonzero. By inserting the completeness condition for $|\{\phi_k\}\rangle$, Eq.~\eqref{eq_corr_general} reads
\begin{widetext}
\begin{align}
&{}_R\langle e^{-i(\theta(x)-\theta(0))}e^{i(A\phi(x)+B\phi(0))}\rangle_R\notag\\
=&\frac{1}{\mathcal N}\int\mathcal D \phi \mathcal D \phi^*\exp\Big\{\sum_{k>0}\Big[-\frac{k}{K^\prime}\Big(e^{\frac{i(\delta_N-\delta_J)}{2}}\Big|\phi_k+\frac{i}{k}\sqrt{\frac{\pi}{L}}e^{-ikx}\Big|^2+ e^{-\frac{i(\delta_N-\delta_J)}{2}}\Big|\phi_k+\frac{i}{k}\sqrt{\frac{\pi}{L}}\Big|^2\Big)\notag\\
&+i\sqrt{\frac{\pi}{L}}\left(A(\phi_k^\prime e^{ikx}+\phi_k^{\prime *} e^{-ikx})+B(\phi_k^\prime +\phi_k^{\prime *})\right)\Big]\Big\}\notag\\
=&\exp\Bigg\{\sum_{k>0}
\Bigg[-\frac{1}{K^\prime\cos((\delta_N-\delta_J)/2)}\frac{\pi}{2kL}(2-e^{ikx}-e^{-ikx})-\frac{K^\prime}{\cos((\delta_N-\delta_J)/2)}\frac{\pi}{2kL}((A+B)^2+2AB(\cos(kx)-1))\notag\\
&+\frac{\pi(e^{ikx}-e^{-ikx})}{kL}\left(\frac{Ae^{-i(\delta_N-\delta_J)/2}-Be^{i(\delta_N-\delta_J)/2}}{2\cos((\delta_N-\delta_J)/2)}-A\right)\Bigg]\Bigg\},
\label{eq_corr_general2}
\end{align}
\end{widetext}
where $\phi_k^\prime = \phi_k + \frac{i}{k}\sqrt{\frac{\pi}{L}}$. The second term in the exponent in Eq.~\eqref{eq_corr_general2} diverges to the negative infinity if $A+B\neq0$. Thus, the expectation value \eqref{eq_corr_general} is nonzero only when $A+B=0$. By substituting $(A$, $B)=(\pm2$, $\mp2)$ in Eq.~\eqref{eq_corr_general}, we obtain
\begin{align}
{}_R\langle e^{-i(\theta(x)-\theta(0))}e^{i(\pm2\phi(x)\mp2\phi(0))}\rangle_R=\left(\frac{\alpha}{x}\right)^{2K_\phi+\frac{1}{2K_\theta}}.\label{eq_expphithetaRR}
\end{align}
Finally, by using Eqs.~\eqref{eq_expthetaRR} and \eqref{eq_expphithetaRR}, we arrive at
\begin{align}
{}_R\langle S^+(x)S^-(0)\rangle_R=C_3\left(\frac{1}{x}\right)^{2K_\phi+\frac{1}{2K_\theta}}+C_4(-1)^x\left(\frac{1}{x}\right)^{\frac{1}{2K_\theta}},\label{eq_S+RR}
\end{align}
where $C_3<0$ and $C_4$ are nonuniversal real correlation amplitudes. We see that the correlation function \eqref{eq_S+RR} is characterized by the anomalous power-law decay with two critical exponents $K_\phi$ and $K_\theta$. Here, we emphasize that as the critical exponents $K_\phi$ and $K_\theta$ are written in terms of the real part of $1/\tilde K$ and that of $\tilde K$, respectively [see Eqs.~\eqref{eq_Kphi} and \eqref{eq_Ktheta}], the universal properties of the right-state spin correlation functions as well as the biorthogonal correlation functions are characterized by the complex-valued TL parameter $\tilde K$.

\section{Exact results}
\label{sec_ExactFinite}
The parameters $\tilde{u}$ and $\tilde{K}$ of the NH TL liquid theory \eqref{eq_Heff0} can be read off from the low-energy spectrum (in the sense of the real part of the energy) of the original lattice Hamiltonian. In this section, we exactly solve the NH XXZ model \eqref{eq_Heff} by using the Bethe ansatz method and obtain the finite-size energy spectrum of the NH XXZ model \eqref{eq_Heff}, demonstrating that the model is described by the complex generalization of the $c=1$ conformal field theory. We also calculate the complex-valued TL parameter from the exchange coupling $J$ and the anisotropy parameter $\Delta_\gamma$. Finally, we discuss the stability conditions to realize the NH TL liquids.

\subsection{Bethe-ansatz solution}
\label{sec_bethe}
%\begin{gather}
%\tilde K =\frac{\pi}{2 \arccos (-\tilde \Delta)},\label{eq_Kexact}\\
%\tilde u = \frac{\pi J}{2\arccos (\tilde \Delta)}\sin (\arccos (\tilde \Delta)),\label{eq_uexact}
%\end{gather}
%where $\tilde \Delta = \tilde J_z / J$. \cite{Fukui98}
We consider the NH XXZ model \eqref{eq_Heff} with length $L$. 
Here we employ the twisted boundary condition $S_{L+1}^+ = e^{i\Phi}S_1^+$ for later convenience. 
%%check: sign of the twist angle
Thanks to the U(1) symmetry of the Hamiltonian, energy eigenstates can be labeled by the number $M$ of down spins. Then, the Bethe equations of the XXZ model are given by \cite{YangYang66_1, YangYang66_2, Takahashi_book}
\begin{equation}
e^{ik_jL-i\Phi}=(-1)^{M-1}\prod_{l\neq j}\frac{1+e^{i(k_j+k_l)}+2\Delta_\gamma e^{ik_j}}{1+e^{i(k_j+k_l)}+2\Delta_\gamma e^{ik_l}},
\label{eq_Bethe1}
\end{equation}
where $k_j\ (j=1,\cdots,M)$ are quasimomenta and $\Delta_\gamma \in\mathbb{C}$. An energy eigenvalue $E_L$ is calculated from the solution of the Bethe equations as
\begin{equation}
E_L(\Phi)=-J\sum_{j=1}^M(\cos k_j+\Delta_\gamma)+\frac{J\Delta_\gamma}{4}L.
\end{equation}
We note that the derivation of the Bethe equations \eqref{eq_Bethe1} still holds for complex $\Delta_\gamma$ and therefore the NH XXZ model \eqref{eq_Heff} is exactly solvable.

The Bethe equations \eqref{eq_Bethe1} are rewritten as
\begin{equation}
\left(\frac{\sinh\frac{\mu}{2}(\lambda_j+i)}{\sinh\frac{\mu}{2}(\lambda_j-i)}\right)^L=e^{i\Phi}\prod_{l\neq j}\frac{\sinh\frac{\mu}{2}(\lambda_j-\lambda_l+2i)}{\sinh\frac{\mu}{2}(\lambda_j-\lambda_l-2i)},
\label{eq_Bethe_TLL}
\end{equation}
where
\begin{gather}
e^{ik_j}=-\frac{\sinh\frac{\mu}{2}(\lambda_j+i)}{\sinh\frac{\mu}{2}(\lambda_j-i)},
\end{gather}
and $\mu=\mathrm{arccos}\Delta_\gamma$. By taking the logarithm of Eq.~\eqref{eq_Bethe_TLL}, we have
\begin{align}
&2L\arctan\left[\frac{\tanh(\mu\lambda_j/2)}{\tan(\mu/2)}\right]\notag\\
=&2\pi I_j+\Phi
+\sum_{l=1}^M 2\arctan\left[\frac{\tanh\frac{\mu}{2}(\lambda_j-\lambda_l)}{\tan\mu}\right],
\label{eq_Bethe_log}
\end{align}
where we set the quantum numbers
\begin{equation}
I_j=-\frac{M+1}{2}+j\ \ (j=1,\cdots,M)
\end{equation}
to obtain the ground state.

\begin{figure}[t]
\includegraphics[width=7.5cm]{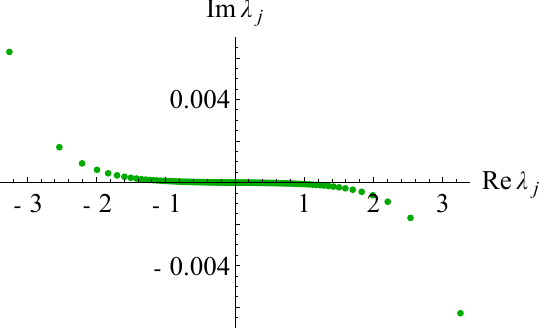}
\caption{Numerical solution of the Bethe equations [Eq.~\eqref{eq_Bethe_log}] for the ground state of the NH XXZ model with $L=2M=250$ and $\Phi=0$. The model parameter is set to $\Delta_\gamma=0.8-0.3i$. Green dots show spin rapidities $\lambda_j\ (j=1,\cdots,M)$.}
\label{fig_Bethe_root}
\end{figure}

In the infinite-size limit with $M/L$ being fixed, the spin rapidities $\lambda_j \ (j=1,\cdots,M)$ are densely distributed along a path $\mathcal{C}$ in the complex plane (see also other cases of NH integrable models \cite{Fukui98, Nakagawa18, Shibata19, Nakagawa21}). In this limit, the Bethe equations \eqref{eq_Bethe_log} reduce to the following integral equation for the distribution function $\sigma_\infty(\lambda)=\lim_{L\to\infty}\sigma_L(\lambda,0)$ of spin rapidities [see Eq.~\eqref{eq_distributionfunction} in Appendix \ref{sec_WienerHopf} for its precise definition]:
\begin{align}
\sigma_\infty(\lambda)=a_1(\lambda)-\int_{\mathcal{C}}d\lambda' a_2(\lambda-\lambda')\sigma_\infty(\lambda'),
\label{eq_Bethe_integ}
\end{align}
where
\begin{equation}
a_n(\lambda)\equiv\frac{1}{2\pi}\frac{\mu\sin(n\mu)}{\cosh(\mu\lambda)-\cos(n\mu)}.
\end{equation}
In Fig.~\ref{fig_Bethe_root}, we show the distribution of spin rapidities obtained from a numerical solution of the Bethe equations \eqref{eq_Bethe_log}. In the Hermitian limit ($\Delta_\gamma\in\mathbb{R}$) with $M/L=1/2$, the spin rapidities are distributed from $-\infty$ to $+\infty$ along the real axis. 
Here, we consider the case in which the path $\mathcal{C}$ can continuously be deformed onto the real axis without crossing the poles of the integrand of Eq.~\eqref{eq_Bethe_integ}. 
%Since the integrand in the right-hand side of Eq.~\eqref{eq_Bethe_integ} exponentially decreases for $|\lambda|\to\infty$, we can deform the path $\mathcal{C}$ onto the real axis without crossing the poles of the integrand. 
Hence, by using the Fourier transformation, we obtain the solution of the integral equation \eqref{eq_Bethe_integ} for $M/L=1/2$ and $\Phi=0$ in the same form of that for the ground state in the Hermitian limit as \cite{Takahashi_book}
\begin{equation}
\sigma_\infty(\lambda)=\frac{1}{4\cosh(\pi\lambda/2)},
\label{eq_sigma_inf}
\end{equation}
and the energy density $e_L(\Phi)\equiv E_L(\Phi)/L$ of the ground state is given by
\begin{align}
e_\infty(0)=&\frac{J\Delta_\gamma}{4}-\frac{2\pi J\sin\mu}{\mu}\int_{\mathcal{C}}d\lambda a_1(\lambda)\sigma_\infty(\lambda)\notag\\
=&\frac{J\Delta_\gamma}{4}-\frac{J\sin\mu}{\mu}\int_{-\infty}^\infty d\omega\frac{\sinh\left(\frac{\pi}{\mu}-1\right)\omega}{2\cosh\omega\sinh(\pi\omega/\mu)}.
\end{align}

The low-energy excitation spectrum is calculated in a similar way. Here we consider a solution with $\Phi=0$ in the $M=L/2-1$ sector and set the quantum numbers as
\begin{align}
I_j=&\frac{L}{4},\frac{L}{4}-1,\cdots,\frac{L}{4}-r+1,\frac{L}{4}-r-1,\notag\\
&\cdots,\frac{L}{4}-s+1,\frac{L}{4}-s-1,\cdots,-\frac{L}{4},
\end{align}
where $0\leq r<s\leq L/4$. Then, the excitation energy $\Delta E$ from the ground state with $M=L/2$ is given by
\begin{equation}
\Delta E=\frac{J\pi\sin\mu}{2\mu}(\sin q_r+\sin q_s),
\label{eq_Bethe_exci}
\end{equation}
where $q_r=2\pi r/L$ and $q_s=2\pi s/L$ are the momenta of spinon excitations. 

To calculate the finite-size energy spectrum of the NH XXZ model, we generalize the Wiener-Hopf method \cite{Takahashi_book} to the NH model. As detailed in Appendix~\ref{sec_WienerHopf}, the leading part of the energy spectrum is given by 
\begin{align}
e_L(\Phi)-e_\infty(0)=-\frac{\pi\tilde{u}}{6L^2}+\frac{2S^2}{\tilde \chi L^2}+\frac{\tilde D_s\Phi^2}{2L^2}+o(1/L^2),
\label{eq_Bethe_finite_result}
\end{align}
where $S=L/2-M$ is the magnetization,
\begin{align}
\tilde{u}%=&v^\prime e^{-i(\delta_N+\delta_J)/2}\notag\\
=&\frac{J\pi\sin\mu}{2\mu}\notag\\
=&\frac{J\pi}{2\arccos\Delta_\gamma}\sin(\arccos\Delta_\gamma)
\label{eq_uexact}
\end{align}
is the complex-valued velocity of excitations,
\begin{equation}
\tilde \chi=\frac{4\mu}{J\pi(\pi-\mu)\sin\mu}
\end{equation}
is the generalized susceptibility, and
\begin{equation}
\tilde D_s=\frac{\pi}{4}\frac{J\sin\mu}{\mu(\pi-\mu)}
\label{eq_Drude_Bethe}
\end{equation}
is the generalized spin stiffness. Equations \eqref{eq_Bethe_finite_result}-\eqref{eq_Drude_Bethe} provide a NH counterpart of the energy spectrum of the XXZ model \cite{Takahashi_book, Hamer85, Hamer86, Hamer87, Woynarovich87, Shastry90, Sutherland90, Tanikawa21}. 

The first term in the right-hand side of Eq.~\eqref{eq_Bethe_finite_result} gives the finite-size correction to the ground-state energy, which is related to the central charge $c=1$ of the conformal field theory \cite{Blote86, Affleck86, Francisco87}. 
This result suggests that a finite-size scaling based on the formula
\begin{equation}
E_L(0)=Le_\infty(0)-\frac{\pi\tilde{u}c}{6L}
\label{eq_Bethe_GS_finite}
\end{equation}
is still valid for the NH model, while the algebraic characterization of the central charge $c$ is nontrivial in the complex extension of the conformal field theory.

\subsection{Finite-size spectrum and the Tomonaga-Luttinger parameter}
\label{sec_finiteCFT}
The finite-size spectrum of the excitation energies in TL liquids was obtained in Refs.~\cite{Haldane81, Haldane81C}. We generalize it to NH TL liquids. We first employ the mode expansion with the bosonization procedure. By rescaling the fields $\phi$ and $\theta$ [see Eqs.~\eqref{eq_phimode} and \eqref{eq_thetamode} in Appendix \ref{sec_bosonization} for their precise definitions] with $K^\prime=\sqrt{v_J/v_N}\in \mathbb R$ and substituting them into the NH TL Hamiltonian \eqref{eq_Heff0}, it is rewritten as
\begin{align}
H_{\mathrm{eff}}^{\mathrm{TL}} = &\frac{v^\prime e^{-i\delta_J}}{4}\sum_{k\neq0}|k|\Big\{(e^{-i(\delta_N-\delta_J)}+1)(b_k^\dag b_{k}+b_{-k} b_{-k}^\dag)\notag\\
&+ (e^{-i(\delta_N-\delta_J)}-1)(b_k^\dag b_{-k}^\dag +b_{-k} b_{k})\Big\}\notag\\
&+ \frac{\pi}{2L}(\tilde v_N(N - N_0)^2 + \tilde v_J J_\mathrm{curt}^2),
\label{eq_mode}
\end{align}
where the zero mode is explicitly written, $v^\prime=\sqrt{v_Jv_N}$, $\tilde v_N = v_Ne^{-i\delta_N}$, $\tilde v_J=v_Je^{-i\delta_J}$, $N_0=L/2$ at half-filling, $N-N_0$ is the change in particle number, and $J_\mathrm{curt}$ is the number of particles that are transferred from the left Fermi point to the right one.
Then, we diagonalize Eq.~\eqref{eq_mode} as follows:
\begin{align}
H_{\mathrm{eff}}^{\mathrm{TL}} = \tilde u \sum_{k\neq0}|k|\bar a_k a_k + \frac{\pi}{2L}(\tilde v_N(N - N_0)^2 + \tilde v_J J_\mathrm{curt}^2).
\label{eq_haldane}
\end{align}
Importantly, the quasiparticle operators $a_k = b_k \cos((\delta_N-\delta_J)/4) -ib_{-k}^\dag \sin((\delta_N-\delta_J)/4)$ and $\bar a_k = b_k^\dag \cos((\delta_N-\delta_J)/4) -ib_{-k} \sin((\delta_N-\delta_J)/4)$ satisfy the commutation relation $[a_k, \bar a_{k^\prime}]=\delta_{k, k^\prime}$ though $\bar a_k \neq a_k^\dag$ \cite{Yamamoto19}. As this transformation is given by the similarity transformation as $a_k = S(\eta)b_k S(-\eta)$, $\bar a_k = S(\eta)b_k^\dag S(-\eta)$ with $S(\eta) =\exp(\frac{i\eta}{2}\sum_{k\neq0}(b_k^\dag b_{-k}^\dag-b_{-k}b_k))$ and $\eta=(\delta_N-\delta_J)/4$, it does not change the conformal-tower structure from the Hermitian limit with the central charge $c=1$. Here, we note that the complex-valued velocities $\tilde u$, $\tilde v_N$, and $\tilde v_J$ are not independent but related by $\tilde u = \sqrt{\tilde v_N \tilde v_J}$, and the first term in Eq.~\eqref{eq_haldane} gives the eigenenergy in Eq.~\eqref{eq_energyharmonic}. Hence, we obtain the excitation energy in a finite system under the periodic boundary condition as
\begin{align}
\Delta E_{\mathrm{PBC}} = \frac{2\pi \tilde u}{L}\left[\frac{1}{4\tilde K}(\Delta N)^2 + \tilde K (\Delta D)^2 + n^+ + n^-\right],
\label{eq_finitePBC}
\end{align}
where $\Delta N = N - N_0\in \mathbb Z$, $\Delta D = J_\mathrm{curt} / 2 \in \mathbb Z$, $n^+$ and $n^-$ are nonnegative integers characterizing particle-hole excitations, and $+(-)$ corresponds to the holomorphic (antiholomorphic) part in conformal field theory. In Eq.~\eqref{eq_finitePBC}, the complex-valued nature of $\tilde K$ and $\tilde u$ gives the complex energy spectrum and leads to dissipation of the system. We see that Eq.~\eqref{eq_finitePBC} is an extension of the finite-size scaling formula in $c=1$ conformal field theory of the Hermitian TL liquids to the NH TL liquids. Equation \eqref{eq_finitePBC} is to be compared with the Bethe-ansatz results [Eq.~\eqref{eq_Bethe_finite_result}] below. From Eq.~\eqref{eq_finitePBC}, we obtain the conformal dimensions as follows:
\begin{align}
\Delta_{\mathrm{CFT}}^\pm = \frac{1}{2}\left(\frac{\Delta N}{2\sqrt {\tilde K}}\pm \Delta D\sqrt {\tilde K}\right)^2 + n^\pm,
\end{align}
which gives the critical exponents of the correlation functions in the infinite system. For NH TL liquids with open boundary conditions (chiral NH TL liquids) \cite{Wen90, Kawakami93}, they do not convey the current, and the excitation energy in a finite system reads
\begin{align}
\Delta E_{\mathrm{OBC}} = \frac{\pi \tilde u}{L}\left(\frac{1}{2\tilde K}(\Delta N)^2 + n \right),
\label{eq_finiteOBC}
\end{align}
where $n$ is a nonnegative integer. Equation~\eqref{eq_finiteOBC} is relevant to the NH-DMRG calculation in Sec.~\ref{sec_NH-DMRG}. The corresponding conformal dimension reads
\begin{align}
\Delta_\mathrm{CFT} =\frac{1}{2\tilde K}(\Delta N)^2 + n.
\end{align}
Thus, the conformal dimensions belong to the universality class characterized by the complex-valued TL parameter $\tilde K$.
\begin{figure}%[t]
\includegraphics[width=7.5cm]{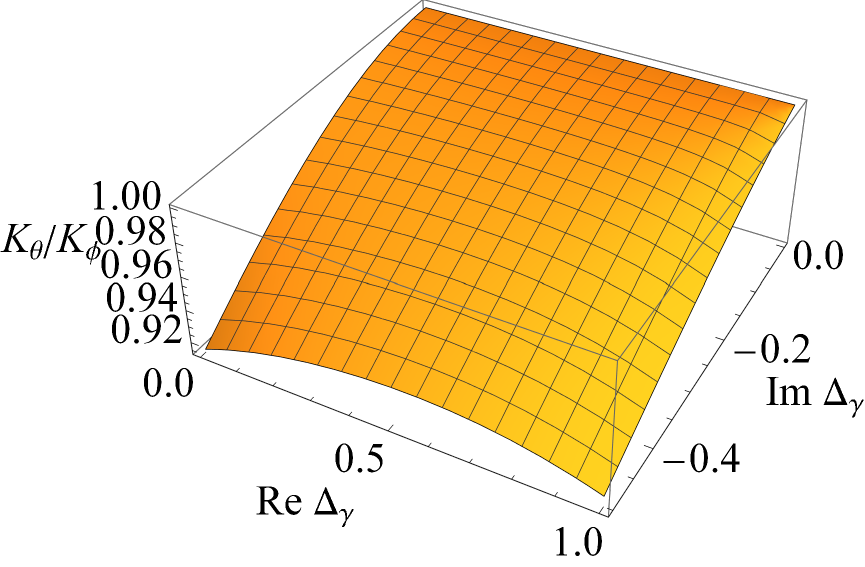}
\caption{Ratio between the critical exponents $K_\phi$ and $K_\theta$ calculated from the exact result \eqref{eq_Kexact} on the NH XXZ model.}
\label{fig_K_ratio}
\end{figure}

We now obtain the TL parameter $\tilde K$. The quantum numbers of the particle number and the current in Eq.~\eqref{eq_finitePBC} are identified as $\Delta N=S$ and $\Delta D=\Phi/(2\pi)$. By comparing Eqs.~\eqref{eq_Bethe_finite_result} and \eqref{eq_finitePBC}, the TL parameter reads
\begin{align}
\tilde{K}=&\frac{\pi}{2(\pi-\arccos\Delta_\gamma)}.
\label{eq_Kexact}
\end{align}
We note that, since the quantum numbers $n^\pm$ of particle-hole excitations in Eq.~\eqref{eq_finitePBC} characterize the spinon excitations in Eq.~\eqref{eq_Bethe_exci}, we obtain $\tilde u$ in Eq.~\eqref{eq_uexact}. In Fig.~\ref{fig_K_ratio}, we plot the ratio between the two critical exponents $K_{\phi}$ and $K_{\theta}$ [see Eqs.~\eqref{eq_Kphi} and \eqref{eq_Ktheta}] calculated from the exact result \eqref{eq_Kexact}. Since the two critical exponents coincide in the Hermitian case, the deviation of the ratio from unity quantifies how the universality class of the NH XXZ model deviates from the standard Hermitian TL liquid theory. 

Next, we discuss how the complex-valued TL parameter $\tilde K$ and the velocity $\tilde u$ of excitations are obtained from numerical calculations of the finite-size spectrum of the excitation energy. As we consider open boundary conditions in the NH-DMRG calculation in Sec.~\ref{sec_NH-DMRG}, we here assume the excitation energy with open boundary conditions given by Eq.~\eqref{eq_finiteOBC}. The parameters $\tilde K$ and $\tilde u$ are calculated through the following two types of energy gaps in a finite system
\begin{gather}
\Delta E_{\mathrm{spectral}} = \Delta E_{\mathrm{OBC}}(\Delta N = 0, n = 1) = \frac{\pi \tilde u}{L},\label{eq_spectral}\\
\Delta E_{\mathrm{spin}} = \Delta E_{\mathrm{OBC}}(\Delta N = 1, n = 0) = \frac{\pi \tilde u}{L}\frac{1}{2\tilde K}\label{eq_spin},
\end{gather}
which we call the spectral gap and the spin gap, respectively. In the Hermitian case, the spectral gap and the spin gap coalesce in the Heisenberg limit. From Eqs.~\eqref{eq_spectral} and \eqref{eq_spin}, we obtain
\begin{gather}
\tilde K =\frac{\Delta E_{\mathrm{spectral}}}{2\Delta E_{\mathrm{spin}}},\label{eq_K}\\
\tilde u = \frac{L\Delta E_{\mathrm{spectral}}}{\pi}.\label{eq_u}
\end{gather}
We use Eqs.~\eqref{eq_K} and \eqref{eq_u} in the NH-DMRG analysis in Sec.~\ref{sec_NH-DMRG}.

\subsection{Stability conditions for realizing the non-Hermitian Tomonaga-Luttinger liquids}
\label{sec_stability}
In contrast to Hermitian TL liquids, the velocity $\tilde u$ of excitations and the TL parameter $\tilde K$ become complex in NH TL liquids. In order for the NH TL liquids to be stable, the real part of the energy spectrum should be bounded from below and the ground state should be normalizable. From Eq.~\eqref{eq_finitePBC}, the energy spectrum is bounded from below when the coefficients satisfy the conditions
\begin{gather}
\mathrm{Re}\left[\frac{\tilde u}{\tilde K}\right]>0,\label{eq_cond1}\\
\mathrm{Re}\big[\tilde u \tilde K\big]>0.\label{eq_cond2}
\end{gather}
We note that the stability condition $\mathrm{Re}[\tilde u]>0$ for the energy spectrum corresponding to the particle-hole excitations is satisfied if the conditions \eqref{eq_cond1} and \eqref{eq_cond2} hold. The conditions \eqref{eq_cond1} and \eqref{eq_cond2} are equivalent to those for ensuring the convergence of the Gaussian integration in the path integrals discussed in Sec.~\ref{sec_correlation}. Moreover, as discussed in Sec.~\ref{sec_correlation}, the condition $-\pi/2<\delta_N-\delta_J<\pi/2$ should be satisfied in order for the ground-state wave function \eqref{eq_wavefunharmonic} to be normalizable. This is equivalent to the following condition:
\begin{align}
\mathrm{Re}\big[\tilde K^2\big]>0.\label{eq_cond3}
\end{align}
As we have obtained the exact solutions for $\tilde u$ and $\tilde K$ in Eqs.~\eqref{eq_uexact} and \eqref{eq_Kexact}, we have to choose appropriate $\Delta_\gamma$ so that the three conditions \eqref{eq_cond1}, \eqref{eq_cond2}, and \eqref{eq_cond3} are satisfied in order to realize the NH TL liquids. If one of Eqs.~\eqref{eq_cond1}, \eqref{eq_cond2}, and \eqref{eq_cond3} is not satisfied, the dissipation causes the instability of the NH TL liquids. The nature of the ground state after the instability is an interesting problem but is left for future studies.

Finally, we explain the physical meaning of the critical exponents in the correlation functions obtained in Sec.~\ref{sec_correlation}. As the operator $e^{i\theta}$ makes a kink in $\phi$, it results in a change in the particle number, thus causing an excitation in the first term in Eq.~\eqref{eq_Heff0} or equivalently in the first term in Eq.~\eqref{eq_finitePBC}. Such an excitation leads to the correlation functions for the field $\theta$ in Eqs.~\eqref{eq_expthetaLR} and \eqref{eq_expthetaRR}. Similarly, as the operator $e^{i\phi}$ creates a current, it causes an excitation in the second term in Eq.~\eqref{eq_Heff0} or equivalently in the second term in Eq.~\eqref{eq_finitePBC}. Such an excitation leads to the correlation functions for the field $\phi$ in Eqs.~\eqref{eq_expphiLR} and \eqref{eq_expphiRR}.

\section{Non-Hermitian density-matrix renormalization group analysis}
\label{sec_NH-DMRG}
We conduct numerical calculations based on the density-matrix renormalization group analysis generalized to non-Hermitian systems (NH-DMRG) \cite{White92, White93, White96, Uri05, Hallberg06, Uri11}. We compare the NH-DMRG results for the lattice Hamiltonian $H_\mathrm{eff}^\mathrm{XXZ}$ [Eq.~\eqref{eq_Heff}] and the analytical results obtained by the effective field theory and the Bethe ansatz discussed in Secs.~\ref{sec_correlation} and \ref{sec_ExactFinite}, thereby demonstrating that the quantum critical phenomena of the model are well described by the NH TL liquid theory.

The algorithm for NH-DMRG is summarized in Appendix \ref{sec_algorithm}. The main idea of NH-DMRG is to use the following form of the density matrix for truncation of eigenstates \cite{Kaulke99}:
\begin{align}
\rho_i = \frac{1}{2}\widehat {\mathrm{Tr}} \left\{|\psi_i\rangle_L{}_L\langle\psi_i| + |\psi_i\rangle_R{}_R\langle\psi_i|\right\},
\label{eq_rhoLR}
\end{align}
where $\widehat {\mathrm{Tr}}$ describes the partial trace on the system block in the DMRG sweep, and $|\psi_i\rangle_{R(L)}$ denotes the right (left) eigenvector corresponding to the $i$th eigenvalue of the NH XXZ Hamiltonian \eqref{eq_Heff}. This type of density matrix was used in Refs.~\cite{Kondev97, Uri99, Affleck04, Uri04}, and its validity was confirmed numerically by comparing the NH-DMRG results with exact results and analytical calculations. We emphasize that, in NH-DMRG, we do not use the traditional variational ansatz to minimize the energy, but we use the numerically accurate density matrix to truncate the eigenstates. In particular, in Ref.~\cite{Uri99}, Eq.~\eqref{eq_rhoLR} was shown to be superior to the other choices of the density matrix such as $\rho_i = \widehat {\mathrm{Tr}} \left\{|\psi_i\rangle_R{}_R\langle\psi_i|\right\}$. By using Eq.~\eqref{eq_rhoLR}, we can avoid all problems related to the possibility of complex eigenvalues. To test the accuracy of the NH-DMRG algorithms, we have compared the ground-state energy obtained from NH-DMRG with that obtained from the exact diagonalization for the number of kept states $m=40$ and the system size $L=14$. We find the numerical error between them is small enough (the maximum numerical error is of the order of $10^{-9}$) so that it does not affect the results obtained below for NH TL liquids. We note that we use the double-precision floating-point numbers for NH-DMRG and the exact diagonalization. We have also calculated the ground-state energy by using the exact diagonalization with quadruple-precision numbers for system sizes $L=10$, $12$, and $14$, and the numerical error between the double precision and the quadruple precision is small enough (the maximum numerical error is of the order of $10^{-9}$).

Below, we assume open boundary conditions, where the numerical accuracy is in general guaranteed to be better than the case with periodic boundary conditions. We assume $\mathrm{Im}\Delta_\gamma <0$ without loss of generality, because the spectrum for $\mathrm{Im}\Delta_\gamma >0$ is obtained by taking complex conjugation to that for $\mathrm{Im}\Delta_\gamma <0$, and the right-state correlation functions discussed below are not affected by the sign of $\mathrm{Im}\Delta_\gamma$. In the following calculations, we have confirmed that all parameters satisfy the stability conditions \eqref{eq_cond1}, \eqref{eq_cond2}, and \eqref{eq_cond3}.

\begin{figure}[t]
\includegraphics[width=8.5cm]{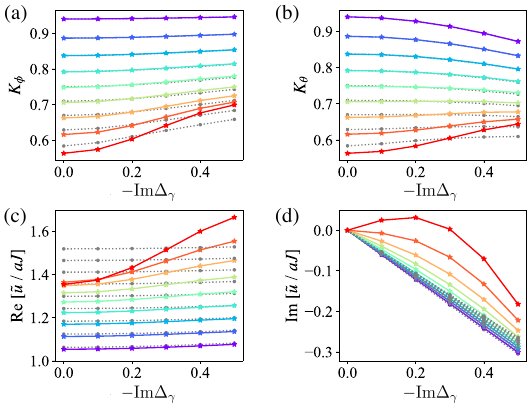}
\caption{Numerical results (solid lines) obtained by the NH-DMRG and exact results (dotted lines) by the Bethe-ansatz solution. (a), (b) Critical exponents $K_\phi$ and $K_\theta$. (c), (d) Velocity of excitations as a function of dissipation $-\mathrm{Im}\Delta_\gamma$. Color plots show the data for $\mathrm{Re}\Delta_\gamma= 0.1, 0.2, 0.3, \cdots, 0.9$ from top to bottom in (a) and (b), and from bottom to top in (c) and (d). The system size is set to $L =130$ and up to 200 states are kept during the NH-DMRG sweep.}
\label{fig_finite_size_scaling}
\end{figure}

By calculating the energy gaps in a finite system (see Appendix~\ref{sec_energygap} for detailed results), we perform a finite-size scaling analysis to extract the critical exponents and the velocity of excitations with Eqs.~\eqref{eq_Kphi}, \eqref{eq_Ktheta}, \eqref{eq_K} and \eqref{eq_u}. The results are shown in Fig.~\ref{fig_finite_size_scaling}, where the exact solutions obtained by Eqs.~\eqref{eq_Kphi}, \eqref{eq_Ktheta}, \eqref{eq_uexact} and \eqref{eq_Kexact} are also shown for comparison. In Fig.~\ref{fig_finite_size_scaling}(a) and (b), we see that the critical exponents $K_\phi$ and $K_\theta$ obtained by NH-DMRG (solid lines) agree quite well with those obtained from the exact solutions (dotted lines) for small $|\Delta_\gamma|$. Though their difference increases for large $|\Delta_\gamma|$, we have conducted further calculations by changing the system size $L$ and conclude that this is a finite-size effect, which may vanish in the thermodynamic limit. For the velocity $\tilde u$ of excitations shown in Figs.~\ref{fig_finite_size_scaling}(c) and (d), the results obtained by NH-DMRG (solid lines) agree well with the exact solutions (dotted lines) for small $|\Delta_\gamma|$. Though their discrepancy becomes singnificant for large $|\Delta_\gamma|$, these features have also been reported in Hermitian cases \cite{Lauchli13, Chen13}. This is because the finite-size effect becomes significant as the model approaches the transition point, at which the cosine term of the NH sine-Gordon model \eqref{eq_NHSG} becomes relevant.

We note that these results are consistent with the renormalization group calculation of the NH sine-Gordon model \cite{Buchhold21}. In Ref.~\cite{Buchhold21}, measurement-induced phase transitions of Dirac fermions were studied in the framework of the replica Keldysh field theory, which gives the NH sine-Gordon Hamiltonian with complex-valued coefficients similar to Eq.~\eqref{eq_NHSG}.  In Ref.~\cite{Buchhold21}, by studying the renormalization group flow of the NH sine-Gordon model, it is pointed out that the model for large $|\tilde{g}_3|$ is governed by the strong-coupling fixed point, where the cosine term in the NH sine-Gordon model becomes relevant. As the finite-size effect is known to become significant when the system approaches the transition point \cite{Lauchli13, Chen13}, the renormalization group analysis is consistent with our NH-DMRG calculation.

It is worth noting that the dependence of $K_\phi$ and $K_\theta$ on dissipation in the NH XXZ model shows interesting behaviors qualitatively different from a model studied before \cite{Ashida16}. In Ref.~\cite{Ashida16}, the effect of dissipation on the critical exponents was studied in the NH Lieb-Liniger model \cite{Lieb86a, Lieb86b}, where the translational invariance ensures that the phase stiffness [$ \tilde u \tilde K$ in our model \eqref{eq_Heff0}] does not change due to dissipation. In the NH Lieb-Liniger model, both critical exponents $K_\phi$ and $K_\theta$ are suppressed by dissipation due to the continuous quantum Zeno effect. Thus, the enhancement of $K_\phi$ and $K_\theta$ in our model [see Fig.~\ref{fig_finite_size_scaling}(a) and (b)] is qualitatively different from the behavior studied in Ref.~\cite{Ashida16}. As this difference is seen in the regime where $|\Delta_\gamma|$ is large and the system approaches the transition point, the Umklapp scattering due to the absence of the continuous translational symmetry in the lattice model affects the behavior of the critical exponents.

\begin{figure}[t]
\includegraphics[width=8.5cm]{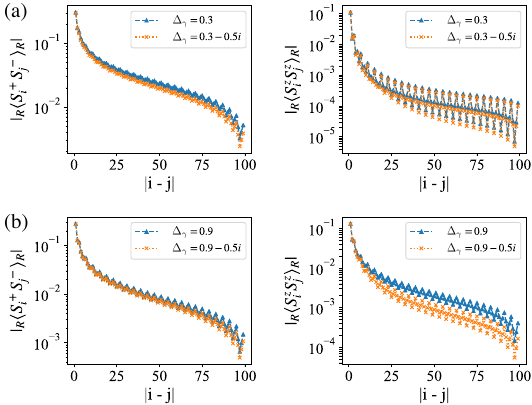}
\caption{Numerical results of correlation functions versus $|i - j|$ for (a) $\mathrm{Re}\Delta_\gamma = 0.3$ and (b) $\mathrm{Re}\Delta_\gamma = 0.9$ obtained by NH-DMRG. The correlation functions are calculated with $i =(L-r+1)/2$ and $j=(L+r+1)/2$ for odd $r$ and $i=(L-r+2)/2$ and $j=(L+r+2)/2$ for even $r$. The system size is set to $L =100$ and up to 170 states are kept during the NH-DMRG sweep.}
\label{fig_correlation_fun}
\end{figure}

Next, by using NH-DMRG, we calculate the experimentally observable correlation functions, that is, the right-state correlation functions \eqref{eq_SzRR} and \eqref{eq_S+RR}. The results are shown in Fig.~\ref{fig_correlation_fun}.
We estimate the maximum numerical error due to truncation to be of the order of $10^{-6}$ for ${}_R\langle S_i^zS_j^z\rangle_R$ and of the order of $10^{-5}$ for ${}_R\langle S_i^+S_j^-\rangle_R$. These numerical errors are of the same orders as those in the Hermitian XXZ model \cite{Hikihara98}, but a much more number of states need to be kept during the NH-DMRG sweep because the convergence problem is severe in the NH case (the number $m$ of kept states is $m=170$ for $L=100$ in the NH case compared with $m=80$ for $L=200$ in the Hermitian case \cite{Hikihara98}). We have also confirmed that the numerical error in the imaginary part of the right-state correlation functions is very small and can be ignored.

In Fig.~\ref{fig_correlation_fun}(a), the $x$-$y$ and $z$ components of the correlation functions with $\mathrm{Re}\Delta_\gamma = 0.3$ are plotted. We see that dissipation suppresses the correlation at long distances compared with that of the Hermitian limit for both ${}_R\langle S_i^zS_j^z\rangle_R$ and ${}_R\langle S_i^+S_j^-\rangle_R$. For $\mathrm{Re}\Delta_\gamma = 0.3$, Fig.~\ref{fig_finite_size_scaling}(a) and (b) show that $K_\phi$ increases with increasing dissipation, and that $K_\theta$ decreases with increasing dissipation. From Eqs.~\eqref{eq_SzRR} and \eqref{eq_S+RR}, these changes in the critical exponents indicate that both correlation functions are suppressed at long distances by dissipation. This is consistent with the NH-DMRG results shown in Fig.~\ref{fig_correlation_fun}(a). In Fig.~\ref{fig_correlation_fun}(b), the correlation functions for $\mathrm{Re}\Delta_\gamma = 0.9$ obtained by NH-DMRG are displayed. We see that both correlation functions ${}_R\langle S_i^zS_j^z\rangle_R$ and ${}_R\langle S_i^+S_j^-\rangle_R$ are suppressed by dissipation at long distances, and the difference between the NH ($\Delta_\gamma = 0.9 - 0.5i$) and the Hermitian ($\Delta_\gamma = 0.9$) cases is smaller for ${}_R\langle S_i^+S_j^-\rangle_R$ and larger for ${}_R\langle S_i^zS_j^z\rangle_R$ than the results for $\mathrm{Re}\Delta_\gamma = 0.3$. For $\mathrm{Re}\Delta_\gamma = 0.9$, the exact solutions shown in Fig.~\ref{fig_finite_size_scaling}(a) and (b) indicate that both $K_\phi$ and $K_\theta$ increase with increasing dissipation. From the results obtained by the field theory in Eqs.~\eqref{eq_SzRR} and \eqref{eq_S+RR}, the exact solution states that ${}_R\langle S_i^zS_j^z\rangle_R$ is suppressed, but ${}_R\langle S_i^+S_j^-\rangle_R$ is enhanced at long distances by dissipation. For ${}_R\langle S_i^zS_j^z\rangle_R$, this result is consistent with the NH-DMRG results shown in Fig.~\ref{fig_correlation_fun}(b), but the behavior of ${}_R\langle S_i^+S_j^-\rangle_R$ seems to be inconsistent. We attribute this inconsistency to the finite-size effect as the model becomes massive for large $|\Delta_\gamma|$. We also note that the behavior of ${}_R\langle S_i^+S_j^-\rangle_R$ with $\Delta_\gamma = 0.9 - 0.5i$ obtained by NH-DMRG seems to be affected by the exponential decay of the correlation functions in the gapped regime, where the model becomes massive.

\begin{figure}[t]
\includegraphics[width=7.5cm]{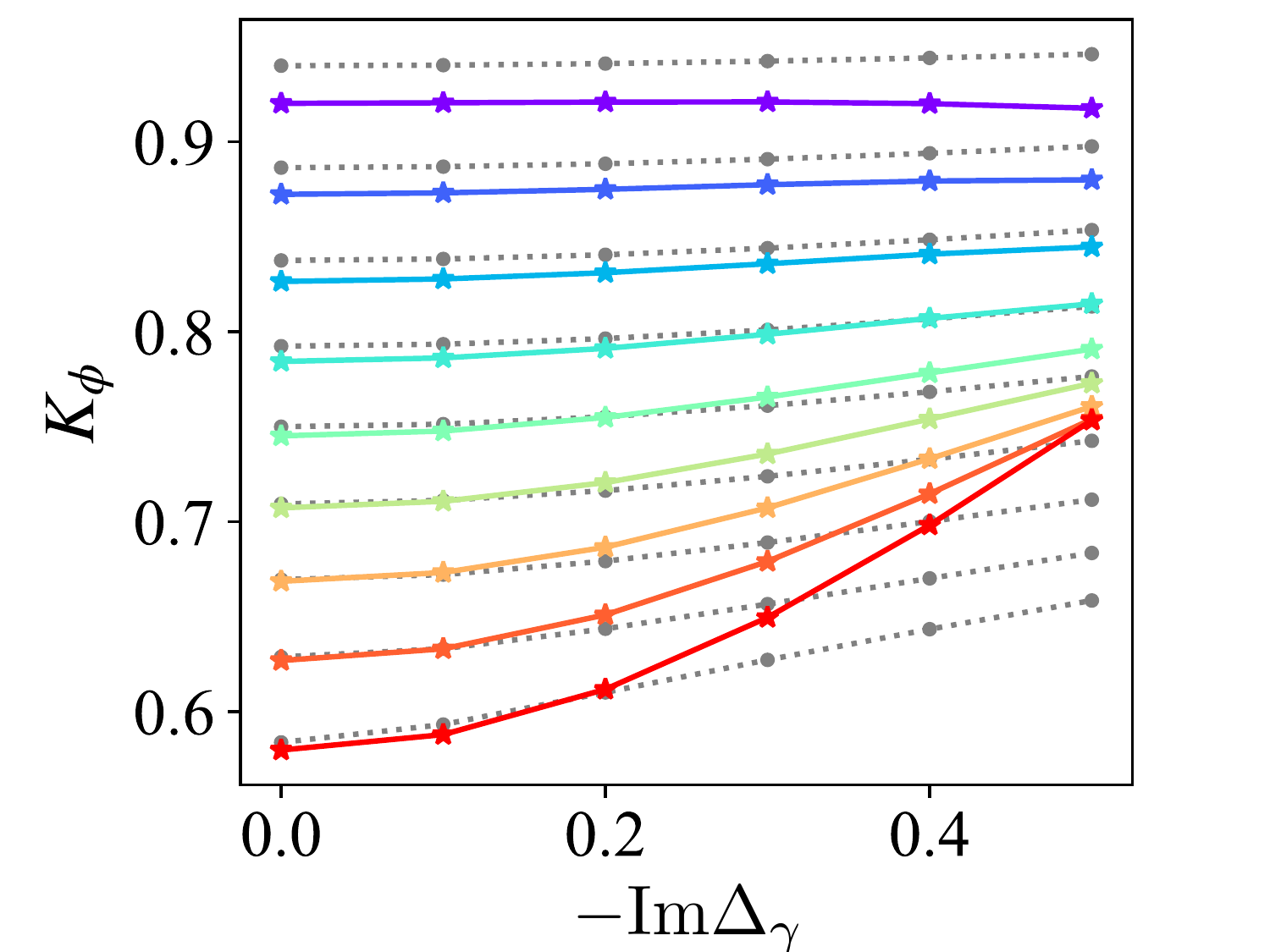}
\caption{Numerical results (solid lines) obtained by the fitting of correlation functions with NH-DMRG and exact results (dotted lines) obtained by the Bethe-ansatz solution of the critical exponent $K_\phi$ as a function of dissipation $-\mathrm{Im}\Delta_\gamma$. Color plots show the data for $\mathrm{Re}\Delta_\gamma = 0.1, 0.2, 0.3, \cdots, 0.9$ from top to bottom. The system size is set to $L =100$ and up to 170 states are kept during the NH-DMRG sweep.}
\label{fig_K_phi_odd_fit}
\end{figure}
To further explore the behavior of the correlation functions obtained by NH-DMRG, we perform a two-parameter fitting for $C_2$ and $K_\phi$ in ${}_R\langle S_i^zS_j^z\rangle_R$ [see Eq.~\eqref{eq_SzRR}]. In NH cases, it has not been explored how the correlation amplitudes and the critical exponents behave in a finite system, and we deal with both the correlation amplitudes and the critical exponents as variables. The details of the fitting procedure are as follows. Since we deal with open boundary conditions, we have removed $11$ sites from both ends of the spin chain (that is, $i=1, 2, \cdots, 11$ and $i=90, 91, \cdots, 100$ for $L=100$) to exclude the effect of the edges as much as possible. We have also removed correlation functions at distances less than $10$ sites (that is, $|i-j|=1, 2, \cdots, 9$) to avoid the higher-order contributions in the correlation functions. We have performed the fitting for the correlation functions at an odd distance and an even distance separately, and confirmed that the parameters obtained by the correlation functions at an odd distance converge faster to those in an infinite system. While we have also performed a fitting for ${}_R\langle S_i^+S_j^-\rangle_R$, there are too many fitting parameters in Eq.~\eqref{eq_S+RR} and therefore the results do not converge with sufficient accuracy. We note that, as the biorthogonal correlation functions and the fitting parameters become complex, it is difficult to perform a fitting for the biorthogonal correlation functions with sufficient accuracy. The critical exponent $K_\phi$ obtained by the fitting of ${}_R\langle S_i^zS_j^z\rangle_R$ at an odd distance is shown in Fig.~\ref{fig_K_phi_odd_fit}. We see that the fitting results show the behavior qualitatively consistent with that of the exact solutions for small $|\Delta_\gamma|$. We have also confirmed that the fitting results for $C_2$ agree qualitatively well with the exact results of an infinite system in the Hermitian limit \cite{Hikihara98} (exact results of $C_2$ for the NH cases are not known). Large difference between the fitting results and the exact results for large $|\Delta_\gamma|$ is due to a finite-size effect as the model approaches the transition point, where the cosine term in the NH sine-Gordon model \eqref{eq_NHSG} becomes relevant. Though the fitting accuracy is low compared to the results obtained by the finite-size scaling analysis in Fig.~\ref{fig_finite_size_scaling}(a) due to the effect of the edges, the results shown in Fig.~\ref{fig_K_phi_odd_fit} give qualitatively the same behavior as those in Fig.~\ref{fig_finite_size_scaling}(a). The NH-DMRG results obtained in this section strongly support the analytical results obtained in Secs.~\ref{sec_correlation} and \ref{sec_ExactFinite}.

%%%-----[Conclusion]-----
\section{Conclusions}
\label{sec_discussion}
We have demonstrated the universal properties of dissipative TL liquids by taking the NH XXZ spin chain as a prototypical model in one-dimensional open quantum many-body systems. First, using the field theory with bosonization, we have calculated two types of correlation functions. Then, we have employed a finite-size scaling approach in conformal field theory and the Bethe ansatz, and obtained exact solutions for the complex-valued TL parameter $\tilde K$ and the velocity $\tilde u$ of excitations. Importantly, we have demonstrated that the model belongs to the universality class characterized by $\tilde K$, which is related to the complex generalization of the $c=1$ conformal field theory. Moreover, we have confirmed that these analytical results are consistent with the numerical results obtained by NH-DMRG for the lattice Hamiltonian in the massless regime with weak dissipation. On the other hand, the NH-DMRG results and the exact solutions start to deviate from each other as the dissipation strength increases. This deviation indicates that the model can be massive for strong dissipation, and this fact is consistent with the renormalization group analysis obtained in Ref.~\cite{Buchhold21}. The NH XXZ spin chain can be derived from the two-component Bose-Hubbard model with two-body loss, which can be realized with long-lived excited states of ytterbium atoms \cite{Spon18, Tomita19}. As a useful tool for controlling dissipation, artificial two-body loss processes with photoassociation techniques \cite{Tomita17} can be utilized to observe the NH TL liquids.

The Bethe ansatz solution of the NH XXZ model for a general complex parameter $\Delta_\gamma$ merits further study. As the integrability of the NH XXZ model holds for arbitrary $\Delta_\gamma\in\mathbb{C}$, it is worthwhile to investigate the possible instability of the NH TL liquid on the basis of exact solutions. Moreover, if the integration path $\mathcal{C}$ in Eq.~\eqref{eq_Bethe_integ} touches a pole of the integrand, the analytic continuation of the solution from the Hermitian limit breaks down and novel criticality may arise at an exceptional point \cite{Nakagawa21}. The exploration of the entire phase diagram of the NH XXZ model remains an interesting research subject.

It is of interest to explore applications of conformal field theory to NH quantum many-body systems. While we have found that the NH TL liquids are successfully described by a complex extension of the $c=1$ conformal field theory, detailed investigation of the Virasoro algebra behind them remains for future studies. Another important question arises concerning a finite-size scaling of entanglement entropy in nonunitary conformal field theories \cite{Bianchini15a, Bianchini15b, Bianchini16, Narayan16, Narayan17, Dupic18}. As a generalization of their frameworks to the biorthogonal bases unique to NH systems has been actively investigated in recent years \cite{Couvreur17, Herviou19, Ryu20}, it is an interesting problem to apply our formalism to the finite-size scaling of entanglement entropy. In view of the fact that conformal field theory in measurement-induced dynamics is investigated in recent years \cite{Matthew19, Ludwig20, Matthew20, Ludwig21}, it is interesting to investigate how the conformal field theory in NH quantum many-body systems is related to the one in the measurement-induced dynamics.

Dissipation breaks the unitarity of the theory and the entire spectrum of the NH XXZ model can be complicated. However, low-energy physics shows universal properties, which lead to unconventional quantum critical phenomena characterized by the complex extension of the $c=1$ conformal field theory. It is our hope that the present paper stimulates further study on dissipative TL liquids in open quantum systems \cite{Buchhold15, Bernier20, Bacsi20, Dora20, Moca21, Bacsi20L, Bacsi21, Dora21}.

\begin{acknowledgments}
We are grateful to Yasuhiko Yamada, Hosho Katsura, and Naoyuki Shibata for fruitful discussions. We particularly thank Hosho Katsura for bringing our attention to the similarity transformation of the quasiparticle operators underlying in Eq.~\eqref{eq_haldane}. This work was supported by KAKENHI (Grants No.\ JP18H01140, No.\ JP18H01145, and No.\ JP19H01838) and a Grant-in-Aid for Scientific Research on Innovative Areas (KAKENHI Grant No.\ JP15H05855) from the Japan Society for the Promotion of Science. K.Y. was supported by WISE Program, MEXT and JSPS KAKENHI Grant-in-Aid for JSPS fellows Grant No.\ JP20J21318. The work of M.T. was supported in part by JSPS KAKENHI (Grants No.\ JP17K17822, No.\ JP20K03787, No.\ JP20H05270, and No.\ JP21H05185). M.N. acknowledges support by KAKENHI (Grant No.\ JP20K14383).
\end{acknowledgments}

\appendix

%%\begin{widetext}
%%\section{Summary of the ground-state property of the Hermitian XXZ model}
%%\label{sec_HermiteXXZ}
%%We give a brief summary of the ground-state property of the conventional XXZ model \eqref{eq_XXZ} for a better understanding of the NH XXZ spin chain. The summary is given in Table~\ref{tab_xxz}.
%%\begin{table*}[h]
%%\caption{Ground state property of the Hermitian XXZ spin chain}
%%\centering
%%\begin{tabular}{c c c c}\hline\hline
%%$\Delta$ & Energy gap & Ground state & Correlation functions \\ \hline
%%\raisebox{0.5em}{$\Delta > 1$} & \raisebox{0.5em}{Ising, gapped} & \raisebox{0.5em}{antiferro, double degeneracy} & \shortstack{antiferromagnetic long-range order in $z$ direction\\ exponential decay in $xy$ plane}\\
%%$\Delta = 1$ & Heisenberg, gapless & no magnetic order, no degeneracy & power law $\times$ log correction\\
%%$\Delta < 1$ & Tomonaga-Luttinger liquid, gapless & no magnetic order, no degeneracy & power law \\ \hline\hline
%%\end{tabular}
%%\label{tab_xxz}
%%\end{table*}
%%\end{widetext}

\section{Bosonization of the NH XXZ model}
\label{sec_bosonization}
We explain the details of the bosonization procedure for the NH XXZ model \eqref{eq_Heff} \cite{Giamarchi03}. Bosonization is conducted in terms of the fields
\begin{align}
\phi(x) = &-(N_R+N_L)\frac{\pi x}{L}\notag\\
&-\frac{i\pi}{L}\sum_{k\neq 0}\left(\frac{L|k|}{2\pi}\right)^{\frac{1}{2}}\frac{1}{k}e^{-\alpha|k|/2-ikx}(b_k^\dag + b_{-k}),
\label{eq_phimode}
\end{align}
and
\begin{align}
\theta(x) = &(N_R-N_L)\frac{\pi x}{L}\notag\\
&+ \frac{i\pi}{L}\sum_{k\neq 0}\left(\frac{L|k|}{2\pi}\right)^{\frac{1}{2}}\frac{1}{|k|}e^{-\alpha|k|/2-ikx}(b_k^\dag - b_{-k}).
\label{eq_thetamode}
\end{align}
These fields obey the commutation relations 
\begin{align}
[\phi(x_1), \theta(x_2)]&=\sum_{k\neq0}\frac{\pi}{Lk}\exp[ik(x_2-x_1)-\alpha|k|]\notag\\
&=i\frac{\pi}{2}\mathrm{sgn}(x_2-x_1),
\end{align}
and 
\begin{align}
[\phi(x_1), \nabla \theta(x_2)]=i\pi\delta(x_2-x_1).
\end{align}
Here, $b_k$ ($b_k^\dag$) is the annihilation (creation) operator of bosons, the subscript $r=R$ ($L$) denotes the right- (left-) going particles, $N_r$ is the number of fermions above the Dirac sea in the $r$ branch, $L$ is the system size, and $\alpha$ is a short-distance cutoff. The fields $\phi(x)$ and $\theta(x)$ are related to the fermion operator as
\begin{align}
\psi_r(x) = &U_r\lim_{\alpha\to 0} \frac{1}{\sqrt{2\pi\alpha}}e^{i(\epsilon_rk_F-\frac{\pi}{L})x}e^{-i(\epsilon_r\phi(x)-\theta(x))},
\end{align}
where $U_r$ is the Klein factor \cite{Giamarchi03}, $\epsilon_R=+1$, and $\epsilon_L=-1$. By using these bosonization dictionaries, we rewrite the NH XXZ Hamiltonian \eqref{eq_Heff} and obtain the NH sine-Gordon Hamiltonian
\begin{align}
H_{\mathrm{eff}}=&H_{\mathrm{eff}}^{\mathrm{TL}}- \frac{2\tilde g_3}{(2\pi\alpha)^2}\int dx \cos(4\phi(x)),
\end{align}
where $\tilde g_3 = a J\Delta_\gamma$, and
\begin{align}
H_{\mathrm{eff}}^{\mathrm{TL}}=\frac{1}{2\pi}\int dx \Big[\tilde u\tilde K(\nabla\theta(x))^2 + \frac{\tilde u}{\tilde K}(\nabla\phi(x))^2\Big],
\end{align}
with
\begin{align}
\tilde u\tilde K&=v_F=J a\sin(k_Fa),\label{eq_uK}\\
\frac{\tilde u}{\tilde K}&=v_F\left(1+\frac{2J\Delta_\gamma a}{\pi v_F}(1-\cos(2k_Fa))\right).
\end{align}
Here, we have assumed the half-filling condition $k_F=\pi/2a$ which corresponds to the zero-magnetization sector of the NH XXZ model. We note that the above derivation is valid only up to the first order of $\Delta_\gamma$. Although Eq.~\eqref{eq_uK} is real up to this order, the imaginary part of $\tilde{u}\tilde{K}$ arises from the higher-order contribution of $\Delta_\gamma$ and cannot be neglected in the NH case.  Thus, we use the exact solution by the Bethe ansatz for the TL parameter $\tilde K$ and the velocity $\tilde u$ of excitations when we compare them with the results obtained from NH-DMRG.

\section{Wiener-Hopf method for a NH integrable model}
\label{sec_WienerHopf}

In this Appendix, we generalize the Wiener-Hopf method \cite{Takahashi_book, Hamer87, Woynarovich87} to calculate the finite-size correction to the ground-state energy of the NH XXZ model. 
A major difference from the Hermitian case is that spin rapidities are not aligned on the real axis but distributed on the complex plane (see Fig.~\ref{fig_Bethe_root}).

We consider the Bethe equations \eqref{eq_Bethe_log} for a large but finite $L$. We define the counting function by
\begin{align}
z_L(\lambda,\Phi)\equiv& -\frac{\Phi}{L}+2\arctan\left[\frac{\tanh(\mu\lambda/2)}{\tan(\mu/2)}\right]\notag\\
&-\frac{1}{L}\sum_{l=1}^M 2\arctan\left[\frac{\tanh\frac{\mu}{2}(\lambda-\lambda_l(\Phi))}{\tan\mu}\right].
\label{eq_Bethe_count1}
\end{align}
The spin rapidities $\lambda_j(\Phi)\ (j=1,\cdots,M)$ are determined from the condition
\begin{equation}
z_L(\lambda_j(\Phi),\Phi)=\frac{2\pi I_j}{L}=\frac{2\pi}{L}\left(-\frac{M+1}{2}+j\right).
\label{eq_Bethe_count2}
\end{equation}
Since $z_L(\lambda,-\Phi)=-z_L(-\lambda,\Phi)$ and $I_{M+1-j}=-I_j$, the spin rapidities satisfy
\begin{equation}
\lambda_j(-\Phi)=-\lambda_{M+1-j}(\Phi),
\end{equation}
and thus the energy eigenvalue
\begin{equation}
E_L(\Phi)=\frac{J\Delta_\gamma}{4}-J\sum_{j=1}^M\frac{\sin^2\mu}{\cosh(\mu\lambda_j(\Phi))-\cos\mu}
\end{equation}
satisfies $E_L(-\Phi)=E_L(\Phi)$.

The distribution function is defined by
\begin{align}
\sigma_L(\lambda,\Phi)\equiv\frac{1}{2\pi}\frac{dz_L(\lambda,\Phi)}{d\lambda},
\label{eq_distributionfunction}
\end{align}
and satisfies
\begin{align}
\sigma_L(\lambda,\Phi)=&a_1(\lambda)-\frac{1}{L}\sum_{j=1}^M a_2(\lambda-\lambda_j(\Phi))\notag\\
=&a_1(\lambda)-\int_{\mathcal{C}'(\Phi)} d\lambda' a_2(\lambda-\lambda')\sigma_L(\lambda',\Phi)\notag\\
&-\int_{\mathcal{C}'(\Phi)} d\lambda' a_2(\lambda-\lambda')S_L(\lambda',\Phi),
\label{eq_Bethe_integ_finite}
\end{align}
where
\begin{equation}
a_n(\lambda)\equiv\frac{1}{2\pi}\frac{\mu\sin(n\mu)}{\cosh(\mu\lambda)-\cos(n\mu)},
\end{equation}
and
\begin{equation}
S_L(\lambda,\Phi)\equiv\frac{1}{L}\sum_{j=1}^M\delta_{\mathcal{C}'}(\lambda-\lambda_j(\Phi))-\sigma_L(\lambda,\Phi).
\end{equation}
Here, we have introduced a path $\mathcal{C}'(\Phi)=\mathcal{C}_+(\Phi)\cup\mathcal{C}(\Phi)\cup\mathcal{C}_-(\Phi)$. The path $\mathcal{C}(\Phi)$ smoothly connects the spin rapidities $\{\lambda_j(\Phi)\}_{j=1}^M$ from $Q_-(\Phi)\equiv\lambda_1(\Phi)$ to $Q_+(\Phi)\equiv\lambda_M(\Phi)$. The endpoints $Q_\pm(\Phi)$ are not located on the real axis in the case of the NH XXZ model. The other two paths $\mathcal{C}_+(\Phi)\equiv\{ Q_+(\Phi)+x;x\in[0,\infty)\}$ and $\mathcal{C}_-(\Phi)\equiv\{ Q_-(\Phi)+x;x\in(-\infty,0]\}$ are half lines parallel to the real axis. The delta function $\delta_{\mathcal{C}'}(\lambda-\lambda_j(\Phi))$ is defined by
\begin{equation}
\int_{\mathcal{C}'(\Phi)} d\lambda F(\lambda)\delta_{\mathcal{C}'}(\lambda-\lambda_j(\Phi))=F(\lambda_j(\Phi))
\end{equation}
for an arbitrary function $F(\lambda)$ defined on $\mathcal{C}'(\Phi)$.

Since the path $\mathcal{C}'(\Phi)$ in the second term on the right-hand side of Eq.~\eqref{eq_Bethe_integ_finite} can be deformed onto the real axis, we have
\begin{align}
&(1+\hat{a}_2(\omega))\hat{\sigma}_L(\omega,\Phi)\notag\\
=&\hat{a}_1(\omega)-\int_{\mathcal{C}'(\Phi)}d\lambda' e^{i\omega\lambda'}\hat{a}_2(\omega)S_L(\lambda',\Phi),
\label{eq_Bethe_finite_Fourier}
\end{align}
where the Fourier transformation of a function $f(\lambda)$ is denoted by
\begin{equation}
\hat{f}(\omega)\equiv \int_{-\infty}^\infty d\lambda f(\lambda)e^{i\omega\lambda}.
\end{equation}
From Eq.~\eqref{eq_Bethe_finite_Fourier}, we have
\begin{align}
\sigma_L(\lambda,\Phi)=\sigma_\infty(\lambda)-\int_{\mathcal{C}'(\Phi)}d\lambda' R(\lambda-\lambda')S_L(\lambda',\Phi),
\label{eq_WH_integ1}
\end{align}
where $\sigma_\infty(\lambda)$ is the distribution function in the infinite system-size limit with $\Phi=0$ [see Eqs.~\eqref{eq_Bethe_integ} and \eqref{eq_sigma_inf}], and
\begin{align}
R(\lambda)\equiv&\frac{1}{2\pi}\int_{-\infty}^\infty d\omega e^{-i\omega\lambda}\frac{\hat{a}_2(\omega)}{1+\hat{a}_2(\omega)}\notag\\
=&\frac{1}{2\pi}\int_{-\infty}^\infty d\omega e^{-i\omega\lambda}\frac{\sinh\left(\frac{\pi}{\mu}-2\right)\omega}{2\sinh\left(\frac{\pi}{\mu}-1\right)\omega\cosh\omega}.
\label{eq_WH_R}
\end{align}

Using the Euler-Maclaurin formula
\begin{align}
&\frac{1}{L}\sum_{j=1}^M F(\lambda_j(\Phi))\notag\\
\simeq&\int_{\mathcal{C}(\Phi)} F(\lambda)\sigma_L(\lambda,\Phi)d\lambda
+\frac{F(Q_+(\Phi))+F(Q_-(\Phi))}{2L}\notag\\
&+\frac{1}{12L^2}\left[\frac{F'(Q_+(\Phi))}{\sigma_L(Q_+(\Phi),\Phi)}-\frac{F'(Q_-(\Phi))}{\sigma_L(Q_-(\Phi),\Phi)}\right],
\label{eq_EulerMaclaurin}
\end{align}
we evaluate the integral in Eq.~\eqref{eq_WH_integ1} and obtain
\begin{align}
\sigma_L(\lambda,\Phi)=&\sigma_\infty(\lambda)+\int_{\mathcal{C}_+(\Phi)\cup\mathcal{C}_-(\Phi)}d\lambda' R(\lambda-\lambda')\sigma_L(\lambda',\Phi)\notag\\
&-\frac{R(\lambda-Q_+(\Phi))+R(\lambda-Q_-(\Phi))}{2L}\notag\\
&+\frac{1}{12L^2}\left[\frac{R'(\lambda-Q_+(\Phi))}{\sigma_L(Q_+(\Phi),\Phi)}-\frac{R'(\lambda-Q_-(\Phi))}{\sigma_L(Q_-(\Phi),\Phi)}\right].
\label{eq_WH_integ_EM}
\end{align}

To apply the Wiener-Hopf method, we introduce the functions 
\begin{align}
g(x,\Phi)&\equiv \sigma_L(x+Q_+(\Phi),\Phi)\notag\\
&=g_+(x,\Phi)+g_-(x,\Phi),
\end{align}
and 
\begin{align}
g_\pm(x,\Phi)\equiv\Theta(\pm x)\sigma_L(x+Q_+(\Phi),\Phi),
\end{align}
for $x\in\mathbb{R}$, where $\Theta(x)$ is the Heaviside unit-step function. Then, using $\sigma_L(\lambda,-\Phi)=\sigma_L(-\lambda,\Phi)$ and $Q_+(-\Phi)=-Q_-(\Phi)$,  
we rewrite Eq.~\eqref{eq_WH_integ_EM} as
\begin{widetext}
\begin{align}
&g(x,\Phi)+\frac{R(x)+R(x+Q_+(\Phi)-Q_-(\Phi))}{2L}-\frac{1}{12L^2}\left[\frac{R'(x)}{\sigma_L(Q_+(\Phi),\Phi)}-\frac{R'(x+Q_+(\Phi)-Q_-(\Phi))}{\sigma_L(Q_-(\Phi),\Phi)}\right]\notag\\
=&\sigma_\infty(x+Q_+(\Phi))+\int_{\mathcal{C}_+(\Phi)\cup\mathcal{C}_-(\Phi)}d\lambda' R(x+Q_+(\Phi)-\lambda')\sigma_L(\lambda',\Phi)\notag\\
=&\sigma_\infty(x+Q_+(\Phi))+\int_0^\infty dx' R(x-x')g(x',\Phi)+\int_0^\infty dx'R(x+x'+Q_+(\Phi)-Q_-(\Phi))\sigma_L(-x'+Q_-(\Phi),\Phi)\notag\\
=&\sigma_\infty(x+Q_+(\Phi))+\int_{-\infty}^\infty dx' R(x-x')g_+(x',\Phi)+\int_{-\infty}^\infty dx' R(x+x'+Q_+(\Phi)-Q_-(\Phi))g_+(x',-\Phi).
\label{eq_WH_integ2}
\end{align}
\end{widetext}
Since $R(x+Q)\simeq -i\mathrm{Res}(\hat{R};-i\pi/2)\exp[-\frac{\pi}{2}(x+Q)]$ for large $\mathrm{Re}[Q]$ ($\mathrm{Res}(f;z_0)$ denotes the residue of a function $f(z)$ at $z=z_0$), the leading part of Eq.~\eqref{eq_WH_integ2} is
\begin{align}
g(x,\Phi)\simeq&\sigma_\infty(x+Q_+(\Phi))+\int_{-\infty}^\infty dx' R(x-x')g_+(x',\Phi)\notag\\
&-\frac{1}{2L}R(x)+\frac{1}{12L^2}\frac{R'(x)}{\sigma_L(Q_+(\Phi), \Phi)},
\end{align}
and its Fourier transformation reads
\begin{align}
\hat{g}_+(\omega,\Phi)+\hat{g}_-(\omega,\Phi)=&\hat{\sigma}_\infty(\omega)e^{-i\omega Q_+(\Phi)}+\hat{R}(\omega)\hat{g}_+(\omega,\Phi)\notag\\
&-\frac{1}{2L}\hat{R}(\omega)-\frac{1}{12L^2}\frac{i\omega \hat{R}(\omega)}{\sigma_L(Q_+(\Phi), \Phi)}.
\label{eq_WH_integ3_Fourier}
\end{align}

Here we use the decomposition
\begin{equation}
1-\hat{R}(\omega)=\frac{1}{G_+(\omega)G_-(\omega)},
\end{equation}
where 
\begin{align}
G_+(\omega)\equiv\frac{\sqrt{2(\pi-\mu)}\Gamma\left(1-\frac{i\omega}{\mu}\right)}{\Gamma\left(\frac{1}{2}-\frac{i\omega}{\pi}\right)\Gamma\left(1-i\omega\frac{\pi-\mu}{\pi\mu}\right)}\left(\frac{\Bigl(\frac{\pi}{\mu}-1\Bigr)^{\frac{\pi}{\mu}-1}}{\Bigl(\frac{\pi}{\mu}\Bigr)^{\frac{\pi}{\mu}}}\right)^{-\frac{i\omega}{\pi}}
\end{align}
[$G_-(\omega)\equiv G_+(-\omega)$] is analytic in the upper (lower) half plane \cite{Takahashi_book, Tanikawa21}. 
%if the imaginary part of Delta is sufficiently small
Then, we have
\begin{align}
&\frac{\hat{g}_+(\omega,\Phi)-C(\omega,\Phi)}{G_+(\omega)}-[G_-(\omega)\hat{\sigma}_\infty(\omega)e^{-i\omega Q_+(\Phi)}]_+\notag\\
=&-G_-(\omega)\hat{g}_-(\omega)+[G_-(\omega)\hat{\sigma}_\infty(\omega)e^{-i\omega Q_+(\Phi)}]_-\notag\\
&-G_-(\omega)C(\omega,\Phi)\notag\\
=&P(\omega,\Phi),
\end{align}
where
\begin{equation}
C(\omega,\Phi)\equiv \frac{1}{2L}+\frac{1}{12L^2}\frac{i\omega}{\sigma_L(Q_+(\Phi), \Phi)},
\end{equation}
$P(\omega,\Phi)$ is an entire function of $\omega$, and
\begin{align}
F(\omega)=&[F(\omega)]_++[F(\omega)]_-,\\
[F(\omega)]_{\pm}\equiv&\frac{\pm i}{2\pi}\int_{-\infty}^\infty d\omega'\frac{F(\omega')}{\omega-\omega'\pm i0}
\end{align}
is the decomposition of $F(\omega)$ into $[F(\omega)]_+$ and $[F(\omega)]_-$ which are analytic in the upper and lower complex planes, respectively \cite{Hamer87}. 
Thus, the solution of Eq.~\eqref{eq_WH_integ3_Fourier} is given by
\begin{align}
\hat{g}_+(\omega,\Phi)=&C(\omega,\Phi)+G_+(\omega)P(\omega,\Phi)\notag\\
&+G_+(\omega)[G_-(\omega)\hat{\sigma}_\infty(\omega)e^{-i\omega Q_+(\Phi)}]_+\notag\\
\simeq&C(\omega,\Phi)+G_+(\omega)P(\omega,\Phi)\notag\\
&+\frac{i}{2}\frac{G_+(\omega)G_-(-i\pi/2)}{\omega+i\pi/2}e^{-\frac{\pi}{2}Q_+(\Phi)},
\label{eq_WH_g_sol}
\end{align}
where we have evaluated the lowest-order contribution to $[G_-(\omega)\hat{\sigma}_\infty(\omega)e^{-i\omega Q_+(\Phi)}]_+$ using the pole $\omega=-i\pi/2$ of $\hat{\sigma}_\infty(\omega)$.

The entire function $P(\omega,\Phi)$ is determined from its asymptotic behavior \cite{Hamer87}. From $\lim_{\omega\to\infty}\hat{g}_+(\omega,\Phi)=0$, $\lim_{\omega\to\infty}\hat{\sigma}_\infty(\omega)=0$, and
\begin{gather}
G_+(\omega)=1+\frac{g_1}{\omega}+\frac{g_1^2}{2\omega^2}+\mathcal{O}\left(\frac{1}{\omega^3}\right),\label{eq_Gplus_asympt}
\end{gather}
we get
\begin{equation}
P(\omega,\Phi)=-C(\omega,\Phi)+\frac{1}{12L^2}\frac{ig_1}{\sigma_L(Q_+(\Phi),\Phi)}.
\label{eq_WH_P}
\end{equation}
From Eqs.~\eqref{eq_WH_g_sol}, \eqref{eq_Gplus_asympt}, and \eqref{eq_WH_P}, the distribution function at the endpoint satisfies
\begin{align}
\sigma_L(Q_+(\Phi),\Phi)=&g(0,\Phi)\notag\\
=&-\lim_{\omega\to\infty}i\omega \hat{g}_+(\omega,\Phi)\notag\\
=&\frac{ig_1}{2L}+\frac{g_1^2}{24L^2\sigma_L(Q_+(\Phi),\Phi)}\notag\\
&+\frac{1}{2}G_-\left(-\frac{i\pi}{2}\right)e^{-\frac{\pi}{2}Q_+(\Phi)}.
\label{eq_WH_endpoint}
\end{align}

From Eq.~\eqref{eq_WH_integ1} and
\begin{align}
\int_{-\infty}^\infty d\lambda a_1(\lambda)R(\lambda-\lambda')=&\int_{-\infty}^\infty d\lambda a_1(\lambda)R(\lambda'-\lambda)\notag\\
=&\frac{1}{2\pi}\int_{-\infty}^\infty d\omega e^{-i\omega\lambda'}\frac{\hat{a}_1(\omega)\hat{a}_2(\omega)}{1+\hat{a}_2(\omega)}\notag\\
=&\int_{-\infty}^\infty d\lambda a_2(\lambda'-\lambda)\sigma_\infty(\lambda)\notag\\
=&a_1(\lambda')-\sigma_\infty(\lambda'),
\end{align}
the difference of the energy density $e_L(\Phi)=E_L(\Phi)/L$ is given by
\begin{align}
\frac{e_L(\Phi)-e_\infty(0)}{A}
=&-\int_{\mathcal{C}'(\Phi)}d\lambda a_1(\lambda)\notag\\
&\times[S_L(\lambda,\Phi)+\sigma_L(\lambda,\Phi)-\sigma_\infty(\lambda)]\notag\\
=&-\int_{\mathcal{C}'(\Phi)}d\lambda a_1(\lambda)\Bigl[S_L(\lambda,\Phi)\notag\\
&-\int_{\mathcal{C}'(\Phi)}d\lambda' R(\lambda-\lambda')S_L(\lambda',\Phi)\Bigr]\notag\\
=&-\int_{\mathcal{C}'(\Phi)}d\lambda\sigma_\infty(\lambda)S_L(\lambda,\Phi),
\end{align}
where $A=2\pi J\sin\mu/\mu$. 
Using the Euler-Maclaurin formula \eqref{eq_EulerMaclaurin} and $\sigma_\infty(\lambda+Q)\simeq\frac{1}{2}\exp[-\frac{\pi}{2}(\lambda+Q)]$ for large $\mathrm{Re}[Q]$, we evaluate the energy difference as
\begin{widetext}
\begin{align}
\frac{e_L(\Phi)-e_\infty(0)}{A}
\simeq&
\int_{-\infty}^\infty dx\sigma_\infty(x+Q_+(\Phi))g_+(x,\Phi)
+\int_{-\infty}^\infty dx\sigma_\infty(x+Q_+(-\Phi))g_+(x,-\Phi)\notag\\
&-\frac{\sigma_\infty(Q_+(\Phi))+\sigma_\infty(Q_-(\Phi))}{2L}
-\frac{1}{12L^2}\left[\frac{\sigma_\infty'(Q_+(\Phi))}{\sigma_L(Q_+(\Phi),\Phi)}-\frac{\sigma_\infty'(Q_-(\Phi))}{\sigma_L(Q_-(\Phi),\Phi)}\right]\notag\\
\simeq&\frac{1}{2}e^{-\frac{\pi}{2}Q_+(\Phi)}\hat{g}_+\left(\frac{i\pi}{2},\Phi\right)
+\frac{1}{2}e^{-\frac{\pi}{2}Q_+(-\Phi)}\hat{g}_+\left(\frac{i\pi}{2},-\Phi\right)\notag\\
&-\frac{1}{4L}(e^{-\frac{\pi}{2}Q_+(\Phi)}+e^{-\frac{\pi}{2}Q_+(-\Phi)})
+\frac{\pi}{48L^2}\left[\frac{e^{-\frac{\pi}{2}Q_+(\Phi)}}{\sigma_L(Q_+(\Phi), \Phi)}+\frac{e^{-\frac{\pi}{2}Q_+(-\Phi)}}{\sigma_L(Q_+(-\Phi),-\Phi)}\right]\notag\\
\simeq&\frac{1}{2}P\left(\frac{i\pi}{2},\Phi\right)G_+\left(\frac{i\pi}{2}\right)e^{-\frac{\pi}{2}Q_+(\Phi)}+\frac{1}{4\pi}\Bigl[G_+\left(\frac{i\pi}{2}\right)e^{-\frac{\pi}{2}Q_+(\Phi)}\Bigr]^2\notag\\
&+\frac{1}{2}P\left(\frac{i\pi}{2},-\Phi\right)G_+\left(\frac{i\pi}{2}\right)e^{-\frac{\pi}{2}Q_+(-\Phi)}+\frac{1}{4\pi}\Bigl[G_+\left(\frac{i\pi}{2}\right)e^{-\frac{\pi}{2}Q_+(-\Phi)}\Bigr]^2.
\label{eq_WH_energy}
\end{align}
\end{widetext}
\begin{table*}[t]
\caption{Algorithm for NH-DMRG}
\centering
\begin{tabular}{l l}\hline \hline
1. & Compute the matrix representation of the spin operators and make the Hamiltonian matrix for four initial blocks (the superblock \\
& consists of a block, two sites, and another block).\\ 
2. & Calculate the right and left ground states of the superblock Hamiltonian by using the Lanzcos method.\\
& The ground state is defined by the lowest real part of the eigenspectrum.\\
3. & Use the right and left ground states $\psi^{R(L)}$ to create the reduced density matrix of the system block as\\
& $\rho_{i_1i_2, i_1^\prime i_2^\prime}=\frac{1}{2}\sum_{i_3 i_4}(\psi_{i_1i_2i_3 i_4}^R\psi_{i_1^\prime i_2^\prime i_3 i_4}^{R*} + \psi_{i_1i_2i_3 i_4}^L\psi_{i_1^\prime i_2^\prime i_3 i_4}^{L*})$.\\
& We emphasize that the validity of this density matrix was confirmed numerically in Ref.~\cite{Uri99},\\
& and we do not use the traditional variational ansatz to minimize the energy.\\
& The density matrix $\rho_{i_1i_2, i_1^\prime i_2^\prime}$ is Hermitian and has real eigenvalues.\\
4. & Diagonalize the density matrix $\rho_{i_1i_2, i_1^\prime i_2^\prime}$ to find a set of eigenvalues and eigenvectors.\\
& Discard all but the largest $m$ eigenvalues and associated eigenvectors.\\
5. & Form a new block by changing the bases to the new $m$ states.\\
6. & Repeat the processes 2-5 to enlarge the system by following the standard infinite-system algorithm.\\
7. & Sweep the superblock by following the standard finite-system algorithm.\\
& Eigenstate prediction is conducted for the right and left eigenstates by the same transformation.\\\hline \hline
\end{tabular}
\label{tab_algorithm}
\end{table*}
On the other hand, since $\mathrm{Re}[\mu]>0$, we have
\begin{align}
\hat{g}_+(0,\Phi)=&\int_{-\infty}^\infty dx g_+(x,\Phi)\notag\\
=&\int_{\mathcal{C}_+(\Phi)}d\lambda\sigma_L(\lambda,\Phi)\notag\\
=&\frac{1}{2\pi}\left[\lim_{x\to\infty}z_L(Q_+(\Phi)+x,\Phi)-z_L(\lambda_M,\Phi)\right]\notag\\
=&\frac{(\pi-\mu)S}{\pi L}+\frac{1}{2L}-\frac{\Phi}{2\pi L},
\label{eq_WH_density}
\end{align}
where $S=L/2-M$ is the magnetization. From Eqs.~\eqref{eq_WH_g_sol} and \eqref{eq_WH_density}, we obtain
\begin{align}
\frac{1}{\pi}G_+\left(\frac{i\pi}{2}\right)e^{-\frac{\pi}{2}Q_+(\Phi)}=&-P(0,\Phi)+\frac{B(\Phi)}{G_+(0)},
\label{eq_WH_density2}
\end{align}
where $B(\Phi)=\frac{(\pi-\mu)S}{\pi L}-\frac{\Phi}{2\pi L}$. Substituting Eq.~\eqref{eq_WH_density2} into Eq.~\eqref{eq_WH_endpoint}, we have
\begin{align}
\sigma_L(Q_+(\Phi),\Phi)=&\frac{2ig_1+\pi}{4L}+\frac{g_1^2-i\pi g_1}{24L^2\sigma_L(Q_+(\Phi),\Phi)}\notag\\
&+\frac{\pi B(\Phi)}{2G_+(0)}.
\label{eq_WH_endpoint2}
\end{align}
Finally, by substituting Eq.~\eqref{eq_WH_density2} into Eq.~\eqref{eq_WH_energy} and using Eq.~\eqref{eq_WH_endpoint2}, we arrive at
\begin{align}
&\frac{e_L(\Phi)-e_\infty(0)}{A}\notag\\
\simeq&-\frac{\pi}{24L^2}+\frac{(\pi-\mu)^2S^2}{2\pi[G_+(0)]^2L^2}+\frac{\Phi^2}{8\pi[G_+(0)]^2L^2}\notag\\
=&-\frac{\pi}{24L^2}+\frac{(\pi-\mu)S^2}{4L^2}+\frac{\Phi^2}{16(\pi-\mu)L^2},
\end{align}
which is equivalent to Eq.~\eqref{eq_Bethe_finite_result} in the main text.

\section{Summary of the NH-DMRG algorithm}
\label{sec_algorithm}
The density-matrix renormalization group (DMRG) analysis is one of the most powerful tools to analyze 1D systems \cite{White92, White93, White96, Uri05, Hallberg06, Uri11}. The principle of DMRG is the variational ansatz, in which the state is optimized by truncating the eigenstates of the density matrix according to the magnitude of the corresponding eigenvalues. However, in NH systems, the variational principle usually breaks down and an important question arises concerning the choice of the density matrix. This is because the right and left eigenvectors of the Hamiltonian are different due to the non-Hermiticity \cite{Kaulke99} and complex eigenvalues can appear in the density matrix. In the previous studies, this problem was tackled by comparing the usage of various types of the density matrix and successful numerical results have been obtained in, e.g., quantum Hall effects \cite{Kondev97}, reaction-diffusion processes \cite{Uri99}, NH TL liquids \cite{Affleck04, Uri04}, and out-of-equilibrium classical systems \cite{Kaulke98, Hieida98}. We note that these situations are essentially different from those of equilibrium quantum systems at a nonzero temperature, where a non-symmetric transfer matrix is generated by applying a Trotter decomposition along the imaginary-time axis \cite{Wang97}. In the latter situations, a similar problem occurs and the non-symmetric density matrix has been used for the truncation of DMRG. However, to the best of our knowledge, no complex eigenvalues stemming from the non-Hermiticity of the transfer matrix have appeared numerically. In this sense, DMRG in NH systems offers intrinsically different situations from those in the equilibrium systems. We apply the NH-DMRG algorithm, in which complex eigenvalues occur as a result of the non-Hermiticity of the Hamiltonian, to NH quantum many-body systems. Such DMRG in NH quantum many-body systems has yet to be fully explored \cite{Affleck04, Uri04, Zhang20}. We use the algorithm detailed in Ref.~\cite{Uri99}, and we give a brief summary of the NH-DMRG algorithm in Table~\ref{tab_algorithm}.

\begin{figure}[b]
\includegraphics[width=8.5cm]{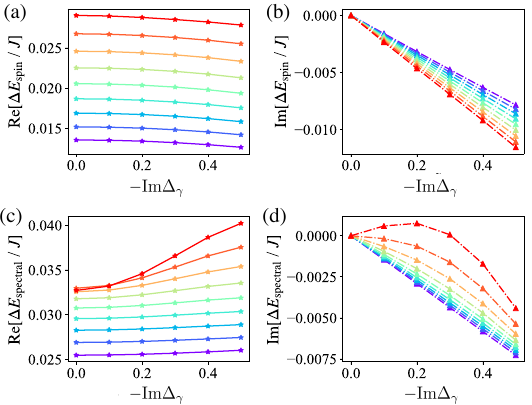}
\caption{NH-DMRG results of energy gaps in a finite system as a function of dissipation $-\mathrm{Im}\Delta_\gamma$. Color plots show the data for $\mathrm{Re}\Delta_\gamma = 0.1, 0.2, 0.3, \cdots, 0.9$ from bottom to top in (a), (c), and (d), and from top to bottom in (b). The system size is set to $L =130$, and up to 200 states are kept during the NH-DMRG sweep.}
\label{fig_energy_gap}
\end{figure}

\section{Detailed results for the energy gaps in a finite system}
\label{sec_energygap}
We show the details of the NH-DMRG results for the energy gap in a finite system of the NH XXZ spin chain. The results are shown in Fig.~\ref{fig_energy_gap}. We estimate the maximum numerical error due to truncation to be of the order of $10^{-8}$ for the ground-state energy. When dissipation is increased, the real part of the spin gap shown in Fig.~\ref{fig_energy_gap}(a) gradually decreases, while the real part of the spectral gap shown in Fig.~\ref{fig_energy_gap}(c) gradually increases. This result shows that the real part of the spectral gap is always larger than that of the spin gap in the weak dissipation regime shown in Fig.~\ref{fig_energy_gap}. As for the imaginary part, the spin gap shown in Fig.~\ref{fig_energy_gap}(b) gradually decreases as dissipation increases, while the imaginary part of the spectral gap shown in Fig.~\ref{fig_energy_gap}(d) seems to increase with increasing dissipation for large $\mathrm{Re}\Delta_\gamma$. To check whether this behavior persists or not in the thermodynamic limit, we have conducted further calculations by changing the system size $L$ (not shown). According to them, we conclude that the increase of $\mathrm{Im}[\Delta E_\mathrm{spectral}/J]$ shown in Fig.~\ref{fig_energy_gap}(d) is a finite-size effect as a result of the higher-order correction to Eqs.~\eqref{eq_spectral} and \eqref{eq_spin} with respect to $1/L$. We observe that the finite-size effect becomes significant for large $|\Delta_\gamma|$.

\section{NH-DMRG results for the gapped regime in the NH XXZ spin chain}
\label{sec_gappedNH-DMRG}
We have also conducted a NH-DMRG calculation for the gapped regime, that is, for $\mathrm{Re}\Delta_\gamma > 1$. However, we have encountered the energy-level crossing problem \cite{Nakamura06} which leads to the breakdown of the NH-DMRG algorithm. We here explain the problem of NH-DMRG in the gapped regime $\mathrm{Re}\Delta_\gamma>1$ in the NH XXZ spin chain. We note that, in this appendix, the gapped regime stands for the regime where the system shows an Ising order with double degeneracy. In the NH-DMRG calculation for the massless regime, the energy levels do not merge with each other as dissipation increases (at least for sufficiently weak dissipation considered in this paper). In this regime, we have not encountered the energy-level crossing problem as we increase the system size during the infinite-system algorithm of NH-DMRG, and the NH-DMRG algorithm works well. However, in the gapped regime, the energy levels are nearly degenerate and cross with each other in the complex plane as dissipation increases \cite{Nakamura06}. One of these points corresponds to exceptional points at which the Hamiltonian cannot be diagonalized. Around such dissipation-induced (exceptional) critical points unique to NH systems, we have found an energy-level crossing problem as we increase the size of the system block during the infinite- and finite-system algorithm of NH-DMRG. In this case, the NH-DMRG algorithm usually breaks down. This is explained as follows. The main procedure in the NH-DMRG algorithm is discarding all but the largest $m$ eigenvalues and associated eigenvectors. In this process, $m$ states sufficient to describe the ground state are kept during NH-DMRG. However, when the energy levels cross and the ground state is changed as we increase the size of the system block, the $m$ kept states no longer describe the ground state of the superblock Hamiltonian in the next iteration. Thus, the $m$ kept states do not describe the ground state with sufficient accuracy and the NH-DMRG algorithm breaks down.

\begin{figure}[t]
\includegraphics[width=7.5cm]{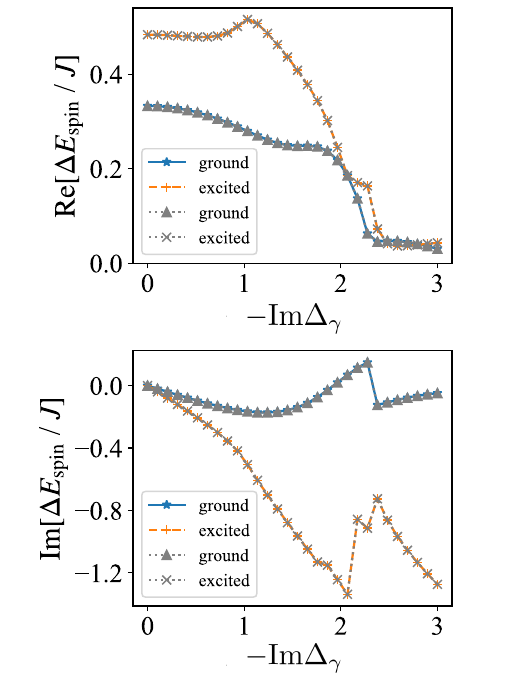}
\caption{Comparison of the energy gaps obtained by NH-DMRG (solid lines and broken lines) and the exact diagonalization (dotted lines). As the ground states for $\mathrm{Re}\Delta_\gamma > 1$ are doubly degenerate in an infinite system, spin gaps for the two lowest states are plotted. The system size is set to $L =14$, $\mathrm{Re}\Delta_\gamma = 1.5$, and up to 40 states are kept during the NH-DMRG sweep. We use the double-precision data for both NH-DMRG and the exact diagonalization (see text).}
\label{fig_gapped}
\end{figure}

In spite of the above problems, we have systematically conducted the NH-DMRG calculation and compared the results with the exact diagonalization by using the double-precision numbers as shown in Fig.~\ref{fig_gapped}. We see that the results obtained by NH-DMRG and the exact diagonalization agree quite well. However, we find that the results are very sensitive to the number of kept states, and in general, the NH-DMRG results are not trustworthy without comparison with those obtained by the exact diagonalization. Moreover, we have performed the exact diagonalization by using quadruple precision for system sizes $L=10$, $12$, and $14$. By comparing the quadruple-precision data and the double-precision data of the ground-state energy obtained from the exact diagonalization, we have found that the numerical error between them increases for large $-\mathrm{Im}\Delta_\gamma$. Thus, we have to pay attention to the precision of the data for large dissipation in the gapped regime. It seems of interest to generalize the NH-DMRG algorithm to gapped regimes in order to explore the ground-state properties. For example, it may be useful to keep all states that are related to the energy-level crossing. This problem is left for future studies.

\nocite{apsrev41Control}
\bibliographystyle{apsrev4-1}
\bibliography{NHXXZ.bib}

\end{document}